\documentclass[11pt, oneside]{article}   	
\usepackage{geometry}
\usepackage{xcolor}              		
\usepackage{ulem}
\usepackage{amsmath}
\usepackage{amsfonts}
\usepackage{setspace}
\usepackage{amssymb}
\usepackage{mathrsfs}
\usepackage{cancel}
\usepackage{slashed}
\usepackage[most]{tcolorbox}
\usepackage{array, bigdelim,multirow,multicol}
\usepackage[toc,page]{appendix}
\usepackage{dsfont}
\usepackage{longtable}
\usepackage{ltablex}
\usepackage{enumitem}
\usepackage{cite}
\usepackage[colorlinks, linkcolor=red, anchorcolor=black, citecolor=blue]{hyperref}
\usepackage[framemethod=tikz]{mdframed}
\usepackage{environ}
\NewEnviron{myalign}{%
\begin{align}
\scalebox{1.}{$\BODY$}
\end{align}
}
\usepackage{graphicx}				
\usepackage{bbm}
\usepackage{bm}								
\newsavebox\myv

\newcommand{\ignore}[1]{}
\newcommand{\dd}{\displaystyle}

\newcommand{\nn}{\nonumber}
\newcommand{\be}{\begin{equation}}
\newcommand{\ee}{\end{equation}}
\newcommand{\bea}{\begin{eqnarray}}
\newcommand{\eea}{\end{eqnarray}}

\newcommand{\TeV}{{\, \rm TeV}}
\newcommand{\GeV}{{\, \rm GeV}}
\newcommand{\MeV}{{\, \rm MeV}}
\newcommand{\keV}{{\, \rm keV}}
\newcommand{\meV}{{\, \rm meV}}

\def\gapmatrix[{\begin{pmatrix}
\gaprows}
\def\gaprows#1[#2]#3{%
\gapcell#2\gapendrow,\ifx]#3\end{pmatrix}\else\afterfi\\\gaprows\fi}

\def\afterfi#1\fi{\fi#1}
\def\gapcell#1,{#1\uppercase{&}\gapcell}
\def\gapendrow#1\gapcell{}

\allowdisplaybreaks

\makeatletter
\renewcommand*{\@fnsymbol}[1]{\ensuremath{\ifcase#1\or *\or  \mathsection\or \ddagger\or
\dagger\or \mathparagraph\or \|\or **\or \dagger\dagger
\or \ddagger\ddagger \else\@ctrerr\fi}}
\makeatother

\makeatletter
\@addtoreset{equation}{section}
\makeatother

\begin{document}
 \unitlength = 1mm

\setlength{\extrarowheight}{0.2 cm}

\title{
\begin{flushright}
{\small TTP24-043,  P3H-24-078}\\
\end{flushright}
\begin{flushright}
\begin{minipage}{0.2\linewidth}
\normalsize 
\end{minipage}
\end{flushright}
 {\Large\bf CPon Dark Matter}\\[0.2cm]}
\date{}

\author{
Ferruccio~Feruglio$^{1}$
\thanks{E-mail: {\tt feruglio@pd.infn.it}}
\ and
Robert Ziegler$^{1,2}$
\thanks{E-mail: {\tt robert.ziegler@kit.edu}}
\
\\*[20pt]
\centerline{
\begin{minipage}{\linewidth}
\begin{center}
$^1${\small
INFN, Sezione di Padova, Via Marzolo~8, I-35131 Padua, Italy}\\
$^2${\small
Institute for Theoretical Particle Physics, KIT, 76128 Karlsruhe, Germany}\\
$^2${\small
Institute for Theoretical Physics, Heidelberg University, 69120 Heidelberg, Germany}\\
\end{center}
\end{minipage}}
\\[8mm]}
\maketitle
\thispagestyle{empty}

\centerline{\large\bf Abstract}
\begin{quote}
\indent
We study a class of supersymmetric models where the strong CP problem is
solved through spontaneous CP violation, carried out by a complex scalar field that determines the Yukawa couplings of the theory. Assuming that one real  component of this field - the CPon - is light, 
we examine the conditions under which it provides a viable Dark Matter candidate.
The CPon couplings to fermions are largely determined by the field-dependent Yukawa interactions, and induce couplings to gauge bosons at 1-loop. All couplings are suppressed by an undetermined UV scale, which needs to  exceed  $10^{12} \GeV$ in order to satisfy constraints on excessive stellar cooling and rare kaon decays. The CPon mass  is limited from below by 5th force experiments and from above by X-ray telescopes looking for CPon decays to photons, leaving a range roughly between 10 meV and 1 MeV. Everywhere in the allowed parameter space the CPon can saturate the observed Dark Matter abundance through an appropriate balance of misalignment and freeze-in production from heavy SM fermions.

\end{quote}
\newpage

\section{Introduction}
The nature of Dark Matter (DM) represents one of the most intriguing problems in contemporary physics~\cite{Cirelli:2024ssz}.
Interpreted in terms of a new still undetected particle, DM can be explained by a variety of possible
candidates, which have been the target of numerous experimental searches. For decades WIMPs have
represented the reigning paradigm, supported by both the naturalness in reproducing the DM abundance and by the belief that low-energy supersymmetry could have offered the solution to the hierarchy problem.
While the WIMP search is still very active today, the failure to detect supersymmetric partners at the LHC and the increasingly severe constraints on the WIMP-nucleon cross-section have gradually turned the attention to other DM candidates.

One of the most appealing possibilities is the QCD axion~\cite{Weinberg:1977ma,Wilczek:1977pj}, designed to solve the strong CP problem~\cite{Peccei:1977hh}.
The axion is the pseudo-Goldstone boson of a global U(1)$_{\rm PQ}$ anomalous symmetry.  
If infrared-dominated by QCD interactions, its dynamics relaxes the physical CP-violating $\bar\theta$ parameter to zero,
whatever amount of CP violation might be stored in the quark mass matrices. In a wide range of parameter space,
the axion is a viable candidate for cold DM, whose abundance is guaranteed, among other possibilities, by the misalignment mechanism~\cite{Preskill:1982cy, Abbott:1982af, Dine:1982ah}.
A robust program of experimental axion searches is currently planned or underway~\cite{DiLuzio:2020wdo}.
A weak point of the axion solution to the strong CP problem is the quality problem, i.e. its excessive sensitivity 
to ultra-violet contributions to the energy density, particularly threatening in the context of a fundamental theory including gravity~\cite{Kamionkowski:1992mf}. Also, the axion solution does not shed any light on the origin of 
fermion masses and mixing angles, unless the U(1)$_{\rm PQ}$ symmetry is embedded in a larger flavor symmetry group explaining Yukawa hierarchies, see e.g. Refs.~\cite{Davidson:1981zd, Wilczek:1982rv, Berezhiani:1989fp, Ema:2016ops, Calibbi:2016hwq}. 

Recently, a new class of solutions to the strong CP problem has been proposed in Refs.~\cite{Feruglio:2023uof,Feruglio:2024ytl}. The ultraviolet theory is assumed to enjoy
CP-invariance, spontaneously broken to deliver a nontrivial CKM
phase while keeping $\bar\theta$ very small. Unlike the Nelson-Barr solution~\cite{Nelson:1983zb,Barr:1984qx},
which also relies on the spontaneous breaking of CP, no additional heavy quark sector is required. Moreover, 
in their minimal implementation, these solutions assume a supersymmetric realization where the field content of the 
Minimal Supersymmetric Standard Model (MSSM) is minimally augmented to include a single extra gauge-invariant chiral supermultiplet $\tau$~\footnote{Another class of solutions of the strong CP problem relying on spontaneous CP violations makes use of discrete symmetries within a multiple Higgs doublet extension of the SM~\cite{Georgi:1978xz,Hall:2024xbd,Ferro-Hernandez:2024snl}. For further works on the solutions to the
strong CP problem based on spontaneous CP violation, see e.g.~\cite{Antusch:2013rla,Vecchi:2014hpa,Dine:2015jga,Schwichtenberg:2018aqc, Valenti:2021rdu,Valenti:2021xjp,Nakagawa:2024ddd}.}. A distinctive feature of the new solutions is that they provide a strict link between the
origin of the fermion mass spectrum and the CP properties of both the weak and strong sectors of the theory.

All physical quantities depend on $\tau$, whose vacuum expectation value (VEV) is the order parameter
for the breaking of an anomaly-free flavour gauge symmetry that incorporates CP.
The strong phase $\bar\theta$ vanishes in the supersymmetric limit, independently on the value of $\tau$
that determines the observed fermion masses, mixing angles and the weak CKM phase.
If the mechanism of supersymmetry breaking does not introduce new physical phases or new flavour patterns,
the observable $\bar\theta$ can remain very small.
Supersymmetric particles have not been detected so far and an interesting possibility that we analyze in this work is that the only low-energy relic of this framework consists of a single
 spin-zero particle, here called CPon (pronounced as {\it cheap-on}), which can provide a viable DM candidate if sufficiently light. 

The CPon shares several features with a CP-violating axion-like particle (ALP), with some important distinctions. Its interactions with all SM particles are non-renormalizable, suppressed by a large UV scale $\Lambda$ taken as a free-parameter. For a sufficiently small mass $m_\xi < \MeV$ and large $\Lambda$, the CPon can only decay into photons and/or neutrinos with a lifetime that easily exceeds
the age of the Universe and can satisfy the stringent constraints on decaying DM from X-ray telescopes.

The CPon can be produced in the early universe by  misalignment  and thermal freeze-in~\cite{Hall:2009bx} in a wide region of the parameter space $(m_\xi,\Lambda)$, similar to anomaly-free ALPs~\cite{Panci:2022wlc, Aghaie:2024jkj}. The allowed region is compact: the CPon mass $m_\xi$  is bounded from below by  precision tests of the gravitational inverse square law, and from above by limits on decaying DM.  The UV scale $\Lambda$ is limited from above by the Planck scale and bounded from below from limits on stellar cooling and SM decays with final state CPons (constraints on CPon-photon conversion in helio- and haloscopes are sub-leading). These limits depend on the CPon-fermion interactions, which are strictly related to the properties of the fermion mass spectrum, with some degree of model-dependence on the level of ${\cal O}(1-100)$ numbers. 

In our analysis we keep the discussion as general as possible. Most of our results rely on few general properties
and, when quantitative implications are derived, they make use of reasonable estimates based on dimensional analysis.
Nevertheless, the most natural realization of the above scenario is within the framework of anomaly-free modular invariant flavour symmetries
\cite{Feruglio:2023uof,Petcov:2024vph,Penedo:2024gtb,Feruglio:2024ytl}.
Modular-invariant scalar potentials can deliver CP-violating minima
\cite{Cvetic:1991qm,Ishiguro:2020nuf,Novichkov:2022wvg,Leedom:2022zdm,Higaki:2024pql}
Moreover, modular invariance, CP-invariance, and field dependence
of observable quantities are all features expected in most 4-dimensional superstring compactifications,
which can also allow for the possibility of a light scalar in the moduli mass spectrum, 
especially for ALP candidates~\cite{Banks:1996ss,Choi:1998dw,Cicoli:2012sz,Cicoli:2021gss,Cicoli:2023opf,Funakoshi:2024yxg}. 
In the final part of this work, we will analyze the prediction of a specific modular and CP invariant model,
whose free parameters are fully determined by fitting fermion masses, mixing angle and the CKM phase.

In the context of modular invariant models, other DM candidates have been proposed, for instance
a light axion~\cite{Higaki:2024jdk,Funakoshi:2024yxg}. Another possible DM candidate is
the lowest mass-state Dirac fermion, a combination of the Weyl components of driving~\footnote{In models with flavor symmetries, driving fields are scalar fields introduced in the scalar potential to achieve the desired vacuum alignment.} and flavon supermultiplets~\cite{Baur:2024lcc}.
Heavy moduli can play the role of DM portal, as discussed in~Ref.~\cite{Chowdhury:2018tzw}. 
In Nelson-Barr solutions to the strong CP problem, ultralight DM candidates have recently been studied
in Ref.~\cite{Dine:2024bxv}. Models unrelated to the strong CP problem,
where a spin-zero component of the field responsible for spontaneous CP violation is a viable DM candidate
have been discussed in Ref.~\cite{Grzadkowski:2018nbc}.

This work is organized as follows. In Section~\ref{S2} we discuss a broad class of models that solve the strong CP Problem by spontaneous CP violation, and introduce the CPon as the light real scalar within the CP-breaking field. Readers who are mainly interested in the CPon couplings to SM field can directly go to Eq.~\eqref{llagr}. In Section~\ref{pheno} we discuss various aspects of CPon phenomenology, in particular constraints  from X-ray telescopes, rare flavor-violating SM decays, long range forces, star cooling and the neutron EDM. In Section~\ref{DM} we discuss CPon  production in the Early Universe by misalignment and freeze-in, and associated constraints from Warm DM. In Section~\ref{QS} we specify two explicit benchmark scenarios, and discuss the resulting parameter space, before concluding in Section~\ref{CL}. In Appendix~\ref{modular} we detail a model with modular-invariance as a complete framework for predicting CPon interactions.

\section{A class of models solving strong CP}
\label{S2}
Our framework consists of a supersymmetric and CP-invariant theory, with the field content of the
 MSSM minimally extended to include a dimensionless gauge-invariant chiral supermultiplet $\tau$~\cite{Feruglio:2023uof,Feruglio:2024ytl}~\footnote{More such multiplets can be present, in general. Here we focus on the most economic realization
where a single multiplet $\tau$ occurs.}.
The theory depends on the K\"ahler potential $K$, a real gauge-invariant function of the fields and their conjugates, the superpotential $w$ and the gauge kinetic functions $f_\alpha$ $(\alpha=1,2,3)$, both gauge-invariant analytic functions of the chiral supermultiplets. All the physical quantities such as masses and coupling constants 
depend on the vacuum expectation value (VEV) of $\tau$.
Under CP, $\tau$ transforms as~\footnote{Up to possible discrete gauge symmetries of the theory acting on $\tau$ non-trivially.  An equivalent formulation makes use of the field $T = -i\tau$, transforming as 
$T \to \bar T$ under CP.} 
\be
\tau\xrightarrow{CP}-\bar\tau\,,
\ee
so that real values of $\tau$ VEV are required to generate the CKM phase, via a spontaneous CP breaking.
Before supersymmetry breaking, the physical angle $\bar\theta$, invariant under colored fermion chiral rotations, is 
\be\nn
\bar\theta=-8\pi^2 {\tt Im}f_3(\tau)+\arg\det Y_U(\tau)Y_D(\tau),
\ee
where $Y_{U,D}(\tau)$ are the matrices of Yukawa couplings in the up and down sectors~\footnote{Notice that $\arg\det \mathscr{M}_U\mathscr{M}_D=\arg\det Y_U(\tau)Y_D(\tau)$, for any K\"ahler potential $K$, $\mathscr{M}_{U,D}$ denoting the quark mass matrices~\cite{Hiller:2001qg,Hiller:2002um}.}. 
It has been shown that, by requiring invariance of the theory under a suitable gauged flavour symmetry, we can achieve the conditions~\cite{Feruglio:2023uof,Feruglio:2024ytl}:
\begin{align}
f_\alpha(\tau)=&~c_\alpha&&(\alpha=1,2,3)\nn\\
\det Y_A(\tau)=&~c_A&&(A=U,D,E)
\label{c2}
\end{align}
where $c_\alpha$ and $c_A$ are constants, required to be real by CP invariance. As a consequence, up to
supersymmetry-breaking contributions, if $c_U c_D>0$ we get $\bar\theta=0$. Assuming a mechanism of supersymmetry breaking that does not generate either new phases or new flavour patterns, at low energy $\bar\theta$ is only corrected by tiny SM contributions~\cite{Ellis:1978hq,Khriplovich:1985jr}
and satisfies the experimental bounds. At the same time, a nontrivial dependence of $Y_{U,D}(\tau)$ on $\tau$ 
can deliver the observed CKM phase. No extra matter multiplets charged under the SM gauge group are needed
and the real part of a single complex spin-zero field $\tau$ is sufficient to spontaneously break CP.

Within this general setup, we consider a scenario where the masses of the superpartners, including the fermionic component of the $\tau$ supermultiplet, are way bigger than the electroweak scale and their effects decouple at
low energies. At the same time, we allow for the possibility that one of the two spin-zero components of
the $\tau$ multiplet remains light and provides a Dark Matter (DM) candidate.
In a general context, the scalar components of the $\tau$ multiplet are typically expected to be heavy, with masses potentially close to the scale $\Lambda$.
However, if the framework under consideration represents (part of) the low-energy limit of a superstring compactification, $\tau$ could correspond to one of its 
moduli~\footnote{Indeed, modular invariance is perhaps the more natural context to 
accommodate the relations in Eq. (\ref{c2})~\cite{Feruglio:2023uof}.}. Moduli masses are suppressed by the gravitino mass $m_{3/2}$ and vanish in the limit of unbroken supersymmetry.
Light moduli are therefore anticipated in scenarios where the gravitino mass is small~\cite{Kusenko:2012ch}, as in the case of gauge mediation, or in situations where the suppression is particularly pronounced. An example of the latter is the large volume scenario, where the mass of the volume modulus is proportional to $m_{3/2}(m_{3/2}/M_P)^{1/2}$~\cite{Quevedo:2014xia}, $M_P$ denoting the Planck mass. In both cases, moduli masses much smaller than one MeV can arise. The lightness of $\tau$ in this context is secured by
supersymmetry, even though the relation between moduli masses and the supersymmetry breaking order parameter $m_{3/2}$ is model-dependent.  

From $f_\alpha= c_\alpha$, we see that there are no tree-level couplings between
$\tau$ and the SM gauge bosons. These can arise from loop corrections.  All the relevant tree-level interactions are of Yukawa type. 
To identify such interactions, we start from the simple case of canonical K\"ahler potential. We expand the Yukawa term around the vacuum $\langle \tau \rangle$,
keeping the first order in the fluctuation $\delta \tau/\Lambda$, $\Lambda$ representing a convenient UV scale.
Working in the limit of massless neutrinos, and neglecting a possible dependence of the Higgsino mass on $\tau$,
we get:
\be
\label{exp}
-\sum_A \psi^c_{Aa}m_{Aab} \psi_{Ab}-\frac{\delta \tau}{\Lambda}\sum_A \psi^c_{Aa}g_{Aab}\psi_{Ab}+...\,,
\ee
where $v_E=v_D$. The sum extends over the three charge sectors ($A=U,D,E)$ and, by denoting derivatives by an index, we have defined
\begin{align}
\label{mandg}
m_{Aab}=&~ \langle Y^A_{ab}(\tau)\rangle  v_A&
g_{Aab}=&~\left\langle Y^A_{\tau ab}(\tau) \right\rangle v_A.
\end{align}
The interactions of the complex field $\delta \tau$ are controlled by the scale $\Lambda$ and by the matrices $g_A$, which, as a consequence of Eq.~\eqref{c2}, satisfy the sum rule:
\be
\label{prg}
{\tt tr}(m_A^{-1} g_A)=0\,.
\ee
This sum rule is independent on the basis chosen for the fermion fields and plays an important role for phenomenology. In the following we further analyze its origin and show that
it holds within a more general class of
K\"ahler potentials. 
\noindent
\subsection{A more general class of models}
Without losing generality and in a matrix notation, the most general K\"ahler potential for quark and lepton
supermultiplets $\varphi_A$ and $\varphi^c_A$ can be parametrized as
\be
\sum_A\bar \varphi^c_{A} \Omega^{c\dagger}_A \Omega^c_A \varphi^c_{A}+\sum_A\bar \varphi_{A} \Omega^\dagger_A \Omega_A\varphi_{A}\,,
\ee
where $\Omega^c_A$ and $\Omega_A$ are matrices that depend on both $\tau$ and $\bar\tau$. By expanding the K\"ahler metric 
around the $\tau$ VEV, we can recover a canonical K\"ahler potential up to terms of second order in the fluctuations $\delta$ and $\bar\delta$ through the transformation
\begin{align}
\label{gocan}
\varphi^c_A\to\langle\Omega_A^{c-1}\rangle\left(\mathbbm{1}-\langle H^c_A\rangle\delta\right)\varphi^c_A~~~~~~~~~~~~~~~~~~~
\varphi_A\to\langle\Omega_A^{-1}\rangle\left(\mathbbm{1}-\langle H_A\rangle\delta\right)\varphi_A\,,
\end{align}
where
\begin{align}
H^c_A=\Omega^c_{A\tau} \Omega_A^{c-1}+\Omega_A^{c\dagger-1}(\Omega^c_{A\bar\tau})^{\dagger}
,~~~~~~~~~~~~~
H_A=\Omega_{A\tau} \Omega_A^{-1}+\Omega_A^{\dagger-1}(\Omega_{A\bar\tau})^\dagger\,.
\end{align}
In the new basis, the mass matrices $m_A$ and couplings $g_A$ of Eq.~\eqref{exp} read
\begin{align}
\label{generalK}
m_{A}=&~ \left\langle\Omega_A^{c-1T}Y^A\Omega^{-1}_A\right\rangle  v_A\,,\nn\\
g_{A}=&~\left\langle\Omega_A^{c-1T}Y^A_\tau\Omega^{-1}_A\right\rangle  v_A-(\langle H^{cT}_A \rangle m_A+m_A \langle H_A\rangle)\,,
\end{align}
When $\Omega_A^c=\Omega_A=\mathbbm{1}$ the K\"ahler potential is canonical and we recover the 
previous case. In the general case, the first contribution to $g_A$ is analogous to the one in Eq.~\eqref{mandg}
and automatically satisfies the sum rule of Eq.~\eqref{prg}. The second contribution is due to the nontrivial
dependence of the K\"ahler metric on $\tau$. It also satisfies Eq.~\eqref{prg}, provided 
\begin{align}
\label{noano}
\det \Omega_A(\tau,\bar\tau)=\det \Omega^c_A(\tau,\bar\tau)=1\,,
\end{align}
which represents the condition for the transformation (\ref{gocan}) to be non-anomalous and to leave the gauge kinetic
functions $f_A$ unaffected. 
Important examples of this more general class are modular invariant theories, which will be further discussed
in Section \ref{MI} and Appendix \ref{modular}. In anomaly-free modular invariant theories, the condition (\ref{noano}) can be naturally satisfied.
\subsection{CPon interactions}
The complex field $\delta \tau$ describes two real mass eigenstates, which are linear combinations of $\delta \tau$ and $\delta \bar \tau$. In the remaining part of this work, we assume
that one of the two mass eigenstates is very heavy, with a mass  of order $\Lambda$, while the other
mass eigenstate, which we denote by $\xi$ and refer to as CPon, has a mass $m_\xi \ll\Lambda$ that we treat as a free parameter.
An angle $\alpha$ (with $0 \le \alpha < \pi$) defines the direction $\xi$ in the $(\delta \tau, \delta \bar \tau)$ plane:
\be
\xi=\frac{1}{\sqrt{2}}(e^{i\alpha}\delta \tau+e^{-i\alpha}\delta \bar \tau)\,.
\ee
After moving to the basis where the fields are canonically normalized, in a four-component notation, the CPon interactions read:
\begin{align}
\label{llagr}
\mathscr{L}_{F}=&~
i \bar \Psi_A\gamma^{\mu}D_\mu \Psi_A-\bar \Psi_A~\hat m_A \Psi_A
-\frac{\xi}{\Lambda} \bar \Psi_A\left(y_S^A+i~ y_P^A\gamma_5\right) \Psi_A \left( 1 + \frac{h}{\sqrt2 v}\right) + \hdots
\end{align}
for  $A=U,D,E$ and the dots stand for terms of order $\xi^2/\Lambda^2$. As we will see, $\Lambda$ is bound to be very large, and terms of order $\xi^2/\Lambda^2$ can be safely neglected.  We also show the CPon couplings to fermions and the Higgs boson $h$, which are relevant for CPon production in the early universe (in our conventions $v = 174 \GeV$). The  fermion couplings are hermitian matrices in flavor space and given by 
\begin{align}
\label{ydef}
y_S^A & = \frac{1}{2\sqrt{2}}(e^{-i\alpha}\hat  g_A+e^{i\alpha}\hat g^\dagger_A) \, , & y_P^A & =\frac{i}{2\sqrt{2}}(e^{-i\alpha}\hat g_A-e^{i\alpha}\hat g^\dagger_A)\,, 
\end{align}
where the hat denotes matrices evaluated in the mass basis~\footnote{The hat denotes the quantities evaluated in the mass basis, reached through
the unitary transformation:
\be\nn
\Psi_A=\left(\begin{array}{c}\psi_A^c\\\psi_A\end{array}\right)\to\left(\begin{array}{cc}U_{A^c} &0\\0&U_A\end{array}\right)\left(\begin{array}{c}\psi_A^c\\\psi_A\end{array}\right)~~~~~~~~~~~(A=U,D,E),
\ee
such that
\be\nn
U_{A^c}^T m_A U_A=\hat m_A~~~~~~~~~~~~~~~~~~~~~~~~U_{A^c}^T g_A U_A=\hat g_A\,,
\ee
where $\hat m$ is diagonal and positive definite.}.
In general, both scalar and pseudoscalar interactions of $\xi$ are present and the CPon behaves as
a CP-violating ALP, with an important distinctive feature that we discuss in the next sections.

\section{CPon Phenomenology}
\label{pheno}
In this section we analyze the possible CPon decay channels and rates, a key aspect to account not only for CPon stability but also for the 
 bounds on decaying DM coming from indirect X-ray and gamma-ray searches.
We also discuss  CPon production through scattering and decay processes initiated by SM particles, which is particularly relevant for determining the CPon abundance in the early universe and assessing constraints from rare flavour-violating decays with the CPon as missing energy. Finally, we analyze the limits on CPon couplings to ordinary matter (nucleons and electrons) from tests of the gravitational inverse square law and star cooling.

\subsection{CPon decays}
\label{lifetime}
For CPon  masses  above threshold, the  decay rate into an electron-positron pair
scales approximately as $\Gamma_{\xi \to ee} \sim m_e^2 m_\xi/8\pi\Lambda^2$. If this is the dominating decay channel, the CPon has a lifetime of about
$10^{22} {\rm \, sec} \times (\Lambda/M_{\rm Pl})^2 \times(10~{\rm MeV}/m_\xi)$. For $\Lambda$ close to the Planck scale the CPon is cosmologically stable, but in conflict with CMB data 
that set a lower bound of about $10^{24}$ sec on the lifetime of a DM candidate decaying into $e^+e^-$, in a mass range from $1$ MeV to $1$ TeV~\cite{Slatyer:2016qyl}. We are thus led to restrict the CPon  mass $m_\xi$ to values smaller than about 1 MeV. In the limit of massless neutrinos, the only decay channel is into two photons~\footnote{ In case of massive neutrinos, the CPon will decay mainly to neutrinos for CPon masses roughly below 1 keV. Still, its total lifetime is larger than $10^{24} {\rm sec}$ for $m_\xi  \le 1 \MeV$ and $\Lambda = 10^{12} \GeV$~\cite{Panci:2022wlc}.}. 

At the one-loop order, the  amplitude for $\xi \to \gamma \gamma$ receives two contributions: one from a loop of electrically charged fermions and one from a loop of their scalar superpartners.
If superpartners are very heavy, the latter can be neglected and we get (in agreement with e.g. Ref.~\cite{Djouadi:2005gi,Djouadi:2005gj,Franceschini:2015kwy}):  
\begin{align}
\label{photondecay}
\Gamma_{\xi\to\gamma\gamma}=\frac{\alpha^2}{1024 \pi^3}\frac{m_\xi^3}{\Lambda^2}\left(|c|^2+|\tilde c|^2\right) \, , 
\end{align}
where~\footnote{In this section we omit the hat symbol to denote fermion masses.}:
\begin{align}
c=&~+\frac{8}{9}\left(4\frac{y_{Suu}}{m_u}+\frac{y_{Sdd}}{m_d}+\frac{y_{Sss}}{m_s}\right)-\frac{1}{3}\left(\frac{y_{Suu}+y_{Sdd}}{m_u+m_d}+\frac{y_{Suu}+y_{Sss}}{m_u+m_s}\right)\nn\\
&~+\frac{28}{81}\left(\frac{y_{Suu}}{m_u}+\frac{y_{Sdd}}{m_d}+\frac{y_{Sss}}{m_s}\right)+{\cal O}\left(m_\xi^2/m_{e,\mu,\tau}^2\right)+{\cal O}\left(m_\xi^2/M_P^2\right)
\label{ctildec}\\
\tilde c=&~-\frac{1}{3}\frac{y_{Pee}}{m_{e}}\frac{m_\xi^2}{m_{e}^2}+{\cal O}\left(m_\xi^2/m_{\mu,\tau}^2\right)+{\cal O}\left(m_\xi^2/M_P^2\right) \,.
\end{align}
Compared to the leading part of the amplitude, the contribution of the scalar superpartners is suppressed by a relative factor of order $(m^2_{t}/m_{\rm SUSY}^2)$, where $m_{\rm SUSY}$ denotes a representative superpartner mass.
The coefficients $c$ and $\tilde c$ describe the CP-conserving and CP-violating part of the decay amplitude, respectively, and exhibit different behaviors. 
Both $c$ and $\tilde c$ are dimensionless and are expected to be proportional to $y_{Saa}/m_a$ and $y_{Paa}/m_a$, at the leading order.
Instead, in both $c$ and $\tilde c$, the leptonic one-loop contribution is suppressed by the ratio $m_\xi^2/m_\ell^2$ compared to the expected leading order behavior. This suppression is a direct consequence of the sum rule in Eq.~\eqref{prg}, and resembles
anomaly-free ALP models~\cite{Takahashi:2020bpq, Han:2020dwo, Han:2022iig,Sakurai:2022roq, Panci:2022wlc}. 
A similar suppression takes place in the hadronic contribution to the CP-violating coefficient $\tilde c$, which is sourced
by a small CPon-$\pi^0$ mixing. Thus, the coefficient $\tilde c$ is dominated by the electron loop, whose result is explicitly shown in Eq.~\eqref{ctildec}.
Finally, the CP-conserving coefficient $c$ receives two distinct hadron contributions:
one from heavy quarks and one from light quarks and gluons. The former can be safely evaluated within the perturbative expansion, by computing the corresponding one-loop diagrams. The latter involves
a non-perturbative regime associated with the light CPon mass. To estimate this part we re-elaborate the results of Refs.~\cite{Leutwyler:1989tn,Flambaum:2024zyt} based on chiral perturbation theory~\footnote{We thank Gabriele Levati for his precious help in deriving the final result for the coefficients $c$ and $\tilde c$.}.
By combining the heavy quark and light meson
contributions, the leading-order term of the CP-conserving coefficient $c$ does not exhibit any particular cancellation.
The first three terms in Eq. (\ref{ctildec}) stand for: 1) the contribution from the heavy quarks $t, c$, and $b$ expressed in terms of
$m_u$, $m_d$ and $m_s$ using the sum rule in Eq.~\eqref{prg}; 2) the contribution from  $\pi^\pm$ and $K^\pm$ loops using chiral perturbation theory~\cite{Flambaum:2024zyt}; 3) the contribution from the $\xi$-gluon-gluon low-energy effective interaction  (from integrating out heavy quarks) using low-energy theorems~\cite{Leutwyler:1989tn} and the sum rule in  Eq.~\eqref{prg}.  As a result, the coefficient $c$ is dominated by the quark sector, and the decay rate of the CPon into two photons is dominated by the CP-conserving part of the amplitude for CPon masses much below 1 MeV.
\subsection{CPon production}
\label{production}
Given the interactions of the CPon with SM fermions in Eq.~\eqref{llagr}, a light CPon can be produced  from SM decays and scatterings, as well as from decays and scatterings of scalar superpartners. The latter processes are suppressed with respect to the former by at least $m_f/m_{\rm SUSY}$, and most relevant for CPon production are flavor-violating decays $f_a \to f_b \xi$ and flavor-conserving $2\to 2$ scattering processes  $f_a \gamma \to f_a \xi$ and $f_a \overline{f}_a \to \gamma \xi$, and their QCD counterparts. While the scattering processes are only relevant for CPon production in the early universe, flavor-violating decays set very stringent bounds on CPon couplings, to be discussed in the next section. Here we collect the relevant expressions for the decay rates and cross-sections. 

\paragraph{Production from SM decays}
From the general Lagrangian of Eq.~\eqref{llagr}  the decay rate for $f_a^A\to f_b^A \xi$ reads\footnote{We omit the $A$ index, which is the same for initial and final states.}  in the limit $m_\xi =0$
\begin{align}
\label{decrate}
\Gamma_{f_a\to f_b\xi} & = \frac{m_a}{16 \pi} \left( 1 - \frac{m_b^2}{m_a^2}\right) \left[  \frac{\left| y_{Sba}\right|^2}{\Lambda^2}  \left(1 + \frac{m_b}{m_a}\right)^2 + \frac{\left| y_{Pba}\right|^2}{\Lambda^2} \left(1 - \frac{m_b}{m_a}\right)^2\right] \, ,
\end{align}
valid for charged lepton decays and flavour-violating transitions among heavy quarks.
Taking also $m_b \ll m_a$, one obtains 
\begin{align}
\label{fdecay}
\Gamma_{f_a \to f_b \xi} & = \frac{m_a}{64 \pi} \frac{\left| g_{ab} \right|^2 + \left| g_{ba} \right|^2}{ \Lambda^2} \, ,
\end{align}
with couplings $g_{ab}$ defined in Eq.~\eqref{generalK}. 
\paragraph{Production from SM $2 \to 2$ scattering processes} 

In two-body scattering processes we only consider the dominant flavor-diagonal channels, with the following cross-sections: 
\begin{align}
\sigma_{f_a \gamma \to f_a \xi}  = &  \frac{\alpha_{\rm em} Q_a^2}{8 s (1-x)^3}  \left[\frac{ y_{Saa}^2}{\Lambda^2} \left( -x^4 + 6 x^3 + 20 x^2 - 22x - 2 (3x+1)^2 \log x - 3 \right) \right. \nonumber \\ & \left. + \frac{ y_{Paa}^2}{\Lambda^2}  (1-x)^2 \left( -x^2 + 4 x - 2 \log x - 3\right)  \right] \, , \\
\sigma_{f_a \overline{f}_a \to  \gamma \xi}  =  & \frac{\alpha_{\rm em} Q_a^2}{s}  \left[ \frac{ y_{Saa}^2}{\Lambda^2}  \left( \frac{4x}{\sqrt{1-4x}} + (1-4x) \tanh^{-1} (\sqrt{1-4x}) \right) \right. \nonumber \\ & \left. +  \frac{ y_{Paa}^2}{\Lambda^2}  \frac{\tanh^{-1} (\sqrt{1-4x})}{1-4x}   \right] \, , 
\end{align}
where $x = m_a^2/s$ and $Q_a$ denotes the electric charge of $f_a$. In the limit of $y_{Saa} = 0$ one recovers the expressions for derivatively coupled axions upon appropriate coupling identification, see e.g. Refs.~\cite{Arias-Aragon:2020shv, Aghaie:2024jkj}.
The corresponding  scattering processes involving gluons and quarks, $\sigma_{q_a g \to q_a \xi}$ and $\sigma_{q_a \overline{q}_a \to g a}$, are obtained from these results by replacing $\alpha_{\rm em} Q_a^2 \to \alpha_s/6$ in $\sigma_{f_a \gamma \to f_a a} $ and  $\alpha_{\rm em} Q_a^2 \to 4 \alpha_s/9$ in $\sigma_{f_a \overline{f}_a \to \gamma a}$. Also relevant is scattering involving Higgs bosons, which in the limit of $\sqrt{s} \gg m_H, m_f$ has the cross-sections 
\begin{align}
\sigma_{f_a h \to f_a \xi}^0 & = \sigma^0_{f_a \overline{f}_a \to h \xi}  = \frac{y_{Saa}^2 + y_{Paa}^2}{64 \pi v^2 \Lambda^2} = \frac{|g_{aa}|^2}{128 \pi v^2  \Lambda^2}\, ,
\end{align}
which are not suppressed in the high-energy limit (as long as $\sqrt{s} \ll \Lambda$) in contrast to the ones involving gauge bosons above. 

Also flavour-violating sfermion decays and scattering lead to CPon production, but their decay rates are suppressed compared to the corresponding processes involving fermions. The ratio of sfermion to fermions  decay rates scales as $m_f/m_{\rm SUSY}$, while the ratio of cross-sections scales as $m_f^2/s$, with $s> m_{\rm SUSY}^2$. We thus neglect the SUSY contribution in the following.

\subsection{Constraints from  CPon decays}
\label{MD}
Electromagnetic decays of DM particles in the keV-MeV range  are constrained by precision measurements of CMB temperature and polarization anisotropies, which give a bound on the lifetime of roughly $\tau_{\gamma\gamma}\gtrsim3\times 10^{24}\text{ sec}$ in the mass range of interest~\cite{Slatyer:2016qyl, Bolliet:2020ofj}. Even stronger constraints arise from searches for X-ray and low-energy gamma lines, which at present give limits on the partial width that are roughly three orders of magnitude more stringent than the CMB, depending on the precise mass range  (with a weak dependence on the DM density profile~\cite{Cirelli:2012ut, Laha:2020ivk}). Here we use the results collected in Appendix A of Ref.~\cite{Panci:2022wlc}, which summarizes  searches that have been conducted  with Chandra~\cite{Watson:2011dw, Horiuchi:2013noa}, Newton-XMM~\cite{Foster:2022ajl}, NuStar~\cite{Perez:2016tcq,Roach:2019ctw,Ng:2019gch,Roach:2022lgo}, and INTEGRAL~\cite{Laha:2020ivk}.  These constraints are expected to further strenghten with future X-ray telescopes, and we use the optimistic projections collected in Ref.~\cite{Panci:2022wlc} for GECCO~\cite{Coogan:2021rez}, THESEUS~\cite{Thorpe-Morgan:2020rwc} and Athena~\cite{Neronov:2015kca,Dekker:2021bos,Ando:2021fhj}, which could  probe lifetimes of order $10^{30} {\rm \, sec}$ for masses in the relevant mass range. For smaller masses there are limits from MUSE spectroscopic observations~\cite{Todarello:2023hdk} (few eV), and the Hubble Space Telescope~\cite{Todarello:2024qci} (few tens of eV), which  however for our benchmark models are weaker than limits from 5th experiments and RG stars, respectively. Instead relevant future limits  are expected from exotic energy injection in the 21-cm power spectrum~\cite{Sun:2023acy} for masses between few tens of eV and few keV.

These limits have to be compared to the prediction of the CPon decay rate into photons in Eq.~\eqref{photondecay}, which is expected to be dominated by quark contribution.
The total CPon decay rate into photons is then given by
\begin{align}
\Gamma_{\xi \to \gamma \gamma} & \approx \frac{1}{3.9 \times 10^{26} {\rm sec}} \left( \frac{m_\xi}{\keV} \right)^3  \left( \frac{10^{12} \GeV}{\Lambda} \right)^2 |c|^2 \, , 
\end{align}
where 
\begin{align}
c\approx \frac{x_{uu}}{0.57 \MeV}+\frac{x_{dd}}{4.7 \MeV}+\frac{x_{ss}}{103 \MeV}\,.
\end{align}
The value of $c$ has been estimated using the light quark $\overline{\rm MS}$ masses renormalized at 2 GeV: $m_u=2.16$ MeV, $m_d=4.70$ MeV and $m_s=93.5$ MeV.
This means that present constraints from CMB and  X-ray searches exclude CPon masses above  roughly $1 \keV$ (for UV scales close to the present limits), with some prospects to probe lower masses with future X-ray telescopes  and 21-cm cosmology. 
\subsection{Constraints from  flavor-violating SM decays}
\label{FC}
As shown in Eq.~\eqref{decrate}, flavour-violating CPon couplings can induce rare decays with a CPon in the final state, which are severely constrained by current bounds on decays with missing energy (for a recent overview see Ref.~\cite{Ziegler:2023aoe}). The most important one is the decay $K^+\to\pi^+ \xi$, which is constrained by the NA62 collaboration~\cite{NA62:2021zjw, Goudzovski:2022vbt} at the level of ${\rm BR} (K^+ \to \pi^+ X) \le 5 \times 10^{-11} $ (90\% CL), with $X$ being a massless invisible particle. For the present scenario with a light CPon the predicted rate reads 
\begin{align}
\Gamma_{K^+\to\pi^+ \xi}=\frac{\left|f_+^{K\pi}(0)\right|^2}{16\pi}\frac{|y_{Sds}|^2}{\Lambda^2}\frac{m_K^3}{(m_s-m_d)^2}\left(1-\frac{m_\pi^2}{m_K^2}\right)^3\,,
\end{align}
where $f_+^{K \pi}(0) = 0.9698 (17)$~\cite{Carrasco:2016kpy, FermilabLattice:2018zqv, FlavourLatticeAveragingGroupFLAG:2021npn}, resulting in a branching ratio
\begin{align}
{\rm BR}(K^+ \to \pi^+ \xi ) & \approx 5 \times 10^{-11}\left( \frac{3.6 \times 10^{11} \GeV}{\Lambda}  \frac{|y_{Sds}|}{\sqrt{m_d m_s}}\right)^2  \, .
\end{align}
Here we have normalized $y_{Sds}$ to the natural value expected in simple scenarios, see Section~\ref{QS}. Since this value can also be easily enhanced by a numerical factor as large as ${\cal O}(100)$, we find that typically one needs $\Lambda \gtrsim 10^{11}\div 10^{13}$ GeV in order to be consistent with NA62 searches. This result  is essentially independent of the CPon mass, as long as it is below the experimental resolution of about few MeV. With the full data set NA62 will be sensitive to 2-body  branching ratios of about $10^{-11}$~\cite{MartinCamalich:2020dfe}, which gives a projected limit on the UV scale that is large by roughly factor two.

Another important channel is the decay $\mu^+\to e^+ X$, which has a signature similar to the SM decay, but with 2-body kinematics. To distinguish signal from background one can employ polarized decays, which are sensitive to the ratio of scalar to pseudo-scalar couplings~\cite{Calibbi:2020jvd}. Unless these couplings are aligned to the SM (i.e. have a V-A structure), the most stringent constraint comes from the Iodidio experiment at TRIUMF~\cite{Jodidio:1986mz}, giving a 90\% CL bound ${\rm BR} (\mu^+\to e^+ X) \le 2.5 \times 10^{-6}$ for a massless $X$ boson. For complete alignment the bound would be loosened to ${\rm BR} (\mu^+\to e^+ X) \le 5.8 \times 10^{-5}$, according to searches by the  TWIST collaboration~\cite{TWIST:2014ymv}. Interestingly, these bounds may be further strengthtened in the near future at MEG-II~\cite{Calibbi:2020jvd, Jho:2022snj}, Mu3e~\cite{Knapen:2023zgi} and Mu2e or COMET~\cite{Hill:2023dym}, probing 2-body branching ratios up to $7 \times 10^{-8}$~\cite{Calibbi:2020jvd}.  In our setup, we get from Eq.~\eqref{fdecay} 
\begin{align}
{\rm BR} (\mu^+\to e^+ X)\approx  \frac{3 \pi^2}{G_F^2 m_\mu^4}\frac{|g_{e\mu}|^2 + |g_{\mu e}|^2}{\Lambda^2}\approx 
2.5 \times 10^{-6} \left( \frac{2.6\times 10^{8}}{\Lambda} \frac{g_{e \mu,\mu e}}{ \sqrt{2 m_e m_\mu}}\right)^2 \, , 
\end{align}
where we have neglected the electron mass and restricted for simplicity to couplings not aligned to the SM. We also introduced the shorthand notation $g_{e \mu, \mu e} \equiv \sqrt {|g_{e\mu}|^2 + |g_{\mu e}|^2}$, which we have again normalized to the natural value expected in simple scenarios. Even taking into account large numerical enhancement factors, it is clear that the stringent constraints on $K^+ \to \pi^+ \xi$ prevent large effects in $\mu^+ \to e^+ \xi$ (even at future experimental facilities), unless there is a  pronounced hierarchy between the couplings in the quark and charged lepton sectors. Similar considerations  for other sectors, e.g. flavor-violating $\tau$- or B-meson decays constrained by Belle II~\cite{MartinCamalich:2020dfe, Belle-II:2022heu, Belle-II:2023esi}, show that such processes are also strongly limited by the large value of $\Lambda$ needed to suppress $b \to d$ transitions. Moreover, as we discuss in the next sections, astrophysical constraints yield limits on $\Lambda$ on the same level as NA62, but are somewhat less model-dependent since they involve only flavor-diagonal couplings. Therefore only $K \to \pi$  decays are  relevant, especially  if the corresponding couplings involve large numerical enhancement factors. 
\subsection{Constraints from long range forces}
\label{LRF}
If the CPon is very light and has sufficiently large couplings to ordinary matter, it can mediate long-range forces that
violate the inverse-square law (ISL), or the equivalence principle (EP), or both. The relevant interactions 
are the scalar ones, described by the $y^A_{S}$ terms in Eq.~\eqref{llagr}, inducing spin-independent effects.
For pseudoscalars interaction, described by the $y^A_{P}$ parameters, spin-dependent effects would arise from the exchange of $\xi$ in the non-relativistic limit. Even if the mass of the CPon is very small or exactly zero, it does not mediate a long-range force between unpolarized bodies~\footnote{Limits form experiments with polarized bodies can be found in Ref.~\cite{OHare:2020wah}. They are less costraining than those discussed here.}.

In a system of two static test bodies with masses $m_{1,2}$ at a distance $r$, the deviation from the Newton potential (ISL) are usually parametrized by:
\be
\label{VISL}
\delta V_{ISL}(r)=-\frac{G m_1 m_2}{r}\alpha e^{\dd - r/\lambda}\,.
\ee

\noindent
The exchange of a light CPon gives rise to a modification to the Newton potential:
\be
\label{nucleon}
\delta V(r)=-\frac{y_{SNN}^2}{4\pi \Lambda^2 r} N_1 N_2 A_1 A_2 e^{\dd -m_\xi r}+ \hdots \, , 
\ee
for two test bodies containing $N_{1,2}$ atoms of mass numbers $A_{1,2}$ experiencing a scalar CPon-nucleus interaction
\be
{\cal L} \supset - \frac{\xi}{\Lambda} y_{SNN}  \bar N N\, , 
\ee
and the dots above stand for additional smaller contributions arising from CPon-electron interactions.
By making use of  the identification
\be
\lambda=\frac{1}{m_\xi}\, , \qquad \qquad \alpha=\frac{y_{SNN}^2}{4\pi G  \Lambda^2 u^2}\,,
\ee
with $u= 0.9315$ GeV being the atomic mass unit, we can
obtain corresponding upper bounds on the coupling constant 
$y_{SNN}$ using the experimental limits on $\lambda$ and $\alpha$.  
For our purposes most relevant are the constraints in the mass range $\meV < m_\xi < 10 \, {\rm  eV}$ (or length scales $100 \, \mu {\rm m} < \lambda < 10\,  {\rm nm} $), which have been obtained by test of the ISL, while EP tests are sensitive only to much smaller CPon masses. Here we use the combined limits from Ref.~\cite{OHare:2020wah}, obtained from results using a  torsional oscillator at the Indiana University-Purdue University Indianapolis (``IUPUI")~\cite{Chen:2014oda}, torsion balance tests (``E\"ot-Wash")~\cite{Kapner:2006si, Lee:2020zjt} and torsion pendula at the Huazhong University of Science and Technology (``HUST")~\cite{Yang:2012zzb, Tan:2020vpf, Tu:2007zz, Tan:2016vwu}.    

To convert these limits into constraints in the plane $(m_\xi,\Lambda)$, we need the relation between the scalar CPon-nucleon couplings $y_{SNN}$
and the scalar CPon-quark couplings $y^{U,D}_{S,aa}/\Lambda$ in Eq.~\eqref{llagr}. 
The leading contribution to the CPon-nucleon coupling, arising from heavy quarks via the triangle diagram 
with external gluons, has been evaluated in Ref.~\cite{Shifman:1978zn}, giving:
\be
\label{Shifman}
y_{SNN}=\frac{2m_N}{27}  \sum_{q=c,b,t}\frac{y_{Sqq}}{m_q}\,.
\ee
A more refined estimate, including also the contribution
of the light quarks, is derived in Ref.~\cite{Cheng:2012qr}. It modifies the value of $y_{SNN}$ by approximately 30\% for the benchmark values of $y_{Sqq}$ in Section~\ref{QS}. For an order-of-magnitude evaluation, we use the relation of Eq.~\eqref{Shifman}, as anyway the limits we obtain on $\Lambda$ depend on the model-dependent value of $y_{Sqq}$. Within this approximation, 
protons and neutrons have the same couplings to the CPon. Taking the natural value for $y_{Sqq} \sim m_q$, we find $y_{SNN} \sim 0.2 \GeV$, so that the bounds from ISL tests restrict $\Lambda$ to be above $(10^{19} \div 10^{13}) \GeV$ in the mass range $(10^{-3} \div 1) \, {\rm  eV}$~\cite{OHare:2020wah}. As discussed in Section~\ref{QS}, these limits can easily strenghtened up to two orders of magnitude, to the presence of large numerical enhancement factors. Requiring $\Lambda$ to be below the Planck scale thus means that the CPon has  to be heavier than about $10^{-9} \MeV$. 

In the above discussion we have neglected the contribution to $\delta V(r)$ arising from CPon
exchange between electrons, which is obtained by replacing $y_{SNN}^2 N_1 N_2 A_1 A_2$
with $y_{See}^2 N_1 N_2 Z_1 Z_2$, in Eq.~\eqref{nucleon}, with $Z_{1,2}$ denoting the atomic numbers of the two test bodies. Such a contribution is therefore parametrically suppressed with respect to the 
one in Eq.~\eqref{nucleon} by a factor $(y_{See}/y_{Stt})^2  (m_t/m_N)^2\approx m_e^2/m_N^2$.
\subsection{Constraints from star cooling }
\label{LRF}
For masses much below 1 MeV the CPon is light enough to be thermally produced in stellar plasmas. Unless the couplings to ordinary matter (electrons and nucleons) are sufficiently small, this leads to an excessive energy loss  in the form of long-lived CPons, which is strongly constrained by observations of various stellar systems~\cite{Raffelt:1996wa}. Here we use the constraints obtained in Ref.~\cite{Bottaro:2023gep,Yamamoto:2023zlu} to set limits on scalar couplings to electron and nucleons, due to CPon production in White Dwarfs (WDs), affecting observations of the WD luminosity function. These limits are effective up to  CPon masses
 of about 1 keV, where production becomes Boltzmann-suppressed, and restrict the scalar couplings to electrons to $y_{See}/\Lambda \lesssim 4 \times 10^{-16}$ and nucleons to $y_{SNN}/\Lambda \lesssim 7 \times 10^{-13}$. For natural values of the  couplings, $y_{See} \sim m_e$ and $y_{SNN} \sim 0.2 \GeV$, it is  clear that the WD limits require $\Lambda \gtrsim 10^{12} \GeV$ for CPon masses below $\sim 1 \keV$. Constraints from Red Giants (RGs), first studied in Ref.~\cite{Grifols:1986fc}, extend to slightly higher temperatures~\footnote{Limits from Horizontal Branch stars extend to even larger masses, but are too weak to be relevant in our scenario.} of about $\sim 10 \keV$, and we use the limits~\footnote{These results have been criticized by the authors of Ref.~\cite{Bottaro:2023gep}, arguing that they are based on the assumption of  non-degenerate and non-relativistic electrons, which is not a good approximation. Nevertheless this should affect the limits on the couplings only at the level of  ${\cal O}(1)$ factors.} provided in  Ref.~\cite{Hardy:2016kme}, which constrain scalar couplings to electrons at the level of $y_{See}/\Lambda \lesssim 7 \times 10^{-16}$ and couplings to nucleons at the level $y_{SNN}/\Lambda \lesssim 1 \times 10^{-12}$. Note that
  these  limits are much stronger than those from e.g. helioscopes~\cite{CAST:2017uph}. 
\subsection{Constraints from the neutron EDM}
\label{BT}
As a result of the conditions in Eq.~\eqref{c2}, $\bar\theta$ (and thus the neutron EDM) vanishes
as long as supersymmetry remains unbroken. 
Small enough corrections to $\bar\theta$ from supersymmetry breaking can be guaranteed under appropriate conditions. 
Denoting by $\Lambda_{\rm SUSY}$  the scale at which supersymmetry breaking is mediated to the observable sector, and
by $m_{\rm SUSY}$ the sparticle mass scale, a favorable framework is achieved when $\Lambda\gg\Lambda_{\rm SUSY}\gg m_{\rm SUSY}$~\cite{Hiller:2001qg,Hiller:2002um}.
At the UV scale $\Lambda$ the observable $\bar\theta$ vanishes and it receives no quantum corrections down to $\Lambda_{\rm SUSY}$ as a consequence of  supersymmetric nonrenormalization theorems~\cite{Ellis:1982tk}. 
Assuming that the whole supermultiplet $\tau$ is heavier than $\Lambda_{\rm SUSY}$, it can be integrated out
without modifying the quark masses. 
Corrections below $\Lambda_{\rm SUSY}$, due to RG running
of soft terms and from integrating out sparticles at the scale $m_{\rm SUSY}$ are model-dependent
and can be kept below the observable level by 
assuming that supersymmetry breaking is gauge-mediated~\cite{Giudice:1998bp} or anomaly-mediated~\cite{Randall:1998uk,Giudice:1998xp,Rattazzi:1999qg}.
Finally, corrections due to the CKM phase~\cite{Ellis:1978hq,Khriplovich:1985jr} are known to be negligibly small.

In our scenario, these conditions are altered by the fact that
one real component of the $\tau$ supermultiplet remains light and the remaining ones 
have a mass of order $m_{\rm SUSY}$. Corrections to $\bar\theta$ are then expected by both integrating out the heavy components of $\tau$, namely a real spin-zero particle and a Majorana fermion,
and by loop corrections involving a CPon exchange. All these corrections 
affect  $\bar\theta$ through a shift of the quark mass matrix, $m_q\to m_q+\delta m_q$, which results in the shift  
\be
\label{dbt}
\delta\bar\theta={\tt Im ~tr}(m_q^{-1} \delta m_q)\,.
\ee
The most important shift $\delta m_q$ arises at one-loop and is necessarily quadratic in $1/\Lambda$, since it involves the emission and absorption of a $\tau$ component with couplings $\propto 1/\Lambda$. On dimensional grounds, we expect contributions of the type
\be
\label{dbt2}
\delta\bar\theta\approx \frac{L}{16 \pi^2}\left\{\frac{v^2}{\Lambda^2},\frac{v~ m_{\rm SUSY}}{\Lambda^2},\frac{m_{\rm SUSY}^2}{\Lambda^2}\right\}\,,
\ee
where $L$ denotes a combination of dimensionless coupling constants, mass ratios, and logarithms~\footnote{See, for example, Ref.~\cite{Enguita:2024nuq}.}.
The largest
set of corrections comes from the third term in Eq. (\ref{dbt2}). Upper bounds on the neutron EDM require $\delta\bar\theta \le 10^{-10}$~\cite{Pospelov:1999mv, Abel:2020pzs}, which translate into an upper limit on the sparticle mass scale
\be
m_{\rm SUSY}<\frac{1.3\times 10^8}{\sqrt{L}}\left( \frac{\Lambda}{10^{12}~{\rm GeV}}\right)~{\rm GeV}.
\ee
At the border of the region allowed by stellar cooling, namely $\Lambda\approx10^{12}$ GeV, 
even considering values of $L$ up to $10^6$, 
$m_{\rm SUSY}$ can be as large as $10^5$ GeV, without spoiling the solution
to the strong CP problem. This makes also clear that the $\delta\bar\theta$ contributions $\propto v^2/\Lambda^2$ are entirely harmless, due to the strong suppression of the UV scale $\Lambda$.

\section{Dark matter abundance}
\label{DM}

As discussed in Section~\ref{lifetime}, the CPon has a lifetime that easily exceeds the age of the universe, and thus is a viable DM candidate. Since its couplings to SM particles are extremely suppressed ($\Lambda \gtrsim 10^{12} \GeV$ from star cooling constraints), the CPon is not in thermal contact with the SM thermal in the early universe, and thus CPon freeze-out production via thermal freeze-out is not an option. Instead a CPon abundance can be produced in the early universe via a variety of mechanisms, here we restrict for simplicity to thermal freeze-in, which is suggested by the smallness of CPon couplings to the SM, and vacuum misalignment.

\subsection{Vacuum misalignment}
\label{VM}


Scalar DM fields are generically produced in the early universe through the misalignment mechanism~\cite{Arias:2012az,Blinov:2019rhb, OHare:2024nmr}, which generalizes the classic scenario for production of the QCD axion~\cite{Preskill:1982cy, Abbott:1982af, Dine:1982ah}. The equation of motion of a scalar field in an expanding universe is given by
\begin{align}
\ddot \xi + 3 H (T) \dot{\xi} + m_\xi^2 \xi = 0 \, , 
\end{align}
where we  restricted to a quadratic CPon potential, and $H$ is the 
Hubble parameter. At early times, the CPon field is frozen at some initial value $\xi_0$ that we parametrize in terms of the UV scale $\Lambda$ as $\xi_0 = \Lambda \theta_0$ (note that the real parameter $\theta_0$ is not a periodic variable). 
We expect $\xi_0$ to remain below the cutoff scale of our effective theory, which is of the order of $\Lambda$. Consequently, $\theta_0$ cannot significantly exceed a value of one.
We assume that the CPon is present before inflation, so that the value $\theta_0$ is uniform across the Hubble patch that forms up the observable universe today. As the universe cools, the CPon starts oscillating, which happens around $m_\xi \sim H (T_{\rm osc})$. We take the condition $m_\xi = 1.6 H (T_{\rm osc})$ to determined $T_{\rm osc}$, which provides a good fit to the results of a numerical integration~\cite{Blinov:2019rhb}. The energy stored in these oscillations behaves just as cold DM, so that today's abundance can be obtained from rescaling the energy density at the onset of oscillations $\rho_\xi = 1/2 m_\xi^2 \Lambda^2 \theta_0^2$ by the ratio of entropy densities today and the onset of oscillations $s_0/s(T_{\rm osc})$.

The final CPon abundance depends on the cosmological scenario at the time of oscillations. For sufficiently high reheating temperatures oscillations start during the epoque of radiation domination (RD). With $H_{\rm RD} = T^2/M_{\rm Pl} 1.66 \sqrt{g_*}$ one obtains 
 \begin{align}
     T_{\rm osc}^{\rm RD} & = 6.7 \times 10^5 \GeV \left( \frac{m_\xi}{\keV} \right)^{1/2}  \, ,               \end{align}
where we took $g_*  ( T_{\rm osc}) = 106.75$. The final CPon abundance in RD is then 
\begin{align}
  \Omega_\xi h^2|_{\rm mis}^{\rm RD} & \approx 0.12  \left(\frac{\Lambda \theta_0}{1.1\times 10^{11}\GeV}\right)^2  \left(\frac{m_\xi}{ \keV}\right)^{1/2}  \, ,
  \label{Oh2misRD}
  \end{align}
  and this expressions are valid for $T_R \ge T_{\rm osc}^{\rm RD} $. 
If instead the reheating temperature is smaller than $T_{\rm osc}^{\rm RD}$, we assume a period of early matter domination (EMD), where $H_{\rm EMD} = H_{\rm RD} \times T^2/T_R^2 \sqrt{g_*/g_*(T_R)}$~\cite{Visinelli:2009kt}. In this case CPon oscillations start at  
\begin{align}
      T_{\rm osc}^{\rm EMD} & = 1.2 \times 10^5 \GeV \left( \frac{m_\xi}{\keV} \right)^{1/4} \left( \frac{T_R}{18 \TeV} \right)^{1/2}  \, , 
              \end{align}
     and the relic abundance is independent of the CPon mass and given by
  \begin{align}
     \Omega_\xi h^2|_{\rm mis}^{\rm EMD} & \approx 0.12  \left(\frac{\Lambda \theta_0}{6.7 \times 10^{11}\GeV}\right)^2  \left(\frac{T_R}{18 \TeV}\right)  \, .
       \label{Oh2misEMD}
     \end{align}
In the following we will treat the reheating temperature as a free parameter above 10 MeV, which slightly exceeds the lower limit allowed by Big-Bang-Nucleosynthesis (BBN)~\cite{Kawasaki:2000en, Hannestad:2004px}.

\subsection{Freeze-in}
\label{FI}

Even for very weak  interactions, unable to keep the CPon in equilibrium with the plasma in the early
universe, DM particles will be produced by cumulative decays and scatterings of SM particles in the thermal bath. This mechanism for building up a DM abundance of the observed size  goes under the name of ``Thermal Freeze-in"~\cite{Hall:2009bx}. Below we briefly review the basis formalism that  allows to relate the DM relic abundance to the model parameters, following the original reference~\cite{Hall:2009bx}, see also the appendices in Refs.~\cite{DEramo:2017ecx,Badziak:2024szg}.

\paragraph{Boltzmann equation}
The number density $n_\xi$ of CPons is determined by the integrated Boltzmann equation (see e.g. Ref.~\cite{Cadamuro:2010cz})

 \begin{align}
\frac{dn_\xi}{dt}+3Hn_\xi= \left( n_\xi^{\rm eq}-n_\xi \right)   \sum_i \Gamma_i \, ,
\end{align}
	where $n_\xi^{\rm eq} =  0.122 \, T^3$ is the CPon equilibrium number density, $H$ is the Hubble parameter $H = T^2/M_{\rm Pl} 1.66 \sqrt{g_* (T)}$
with 	$g_* (T)$ denoting the total number of relativistic  degrees of freedom
and $\Gamma_i$ are the specific CPon production rates, which are related to the respective collision terms ${\cal C}_i$ by  $\Gamma_i = {\cal C}_i/n_\xi^{\rm eq}$.

Entropy conservation ($d(sa^3)/dt = 0$) allows to rewrite the time derivative in terms of a derivative with respect to temperature as $dT/dt = - H T$, which is valid when the effective number of relativistic entropy degrees of freedom is approximately constant, $dg_{*s} (T)/dT \approx 0$.  This relation can be used to rewrite the Boltzmann equation in terms of the CPon yield $Y_\xi = n_\xi/s$, giving 
 \begin{align}
\frac{d Y_\xi}{dT} = - \left( 1 -\frac{Y_\xi}{Y_\xi^{\rm eq}} \right)   \sum_i \frac{{\cal C}_i (T)}{s T H} \, ,
\end{align}
with the entropy density $s = 0.439 \, T^3 g_{*s} (T)$. In the freeze-in regime the CPons are never in thermal equilibrium, $Y_\xi \ll Y_\xi^{\rm eq}$, and their initial abundance at $T_R$ can be neglected. Thus  the final yield $Y^0_\xi$ of CPons today (at $T \approx 0$) is given by the integral
\begin{align}
Y^0_\xi =  \sum_i \int_0^{T_R} \frac{  {\cal C}_i (T)}{s T H} dT = \sum_i  \frac{1.4}{g_{*s} \sqrt{g_*}} \int_0^{T_R} \frac{{\cal C}_i (T) M_{\rm Pl}}{T^6} dT  \, ,
\end{align}
where the effective number of relativistic degrees of freedom are evaluated at the characteristic temperature of the production process, which for decays is the mass of the decaying particle and for scattering processes the threshold center-of-mass energy (unless the process is UV sensitive, in which case it is $T_{\rm max} = T_R$~\cite{Hall:2009bx, Elahi:2014fsa}).  The final CPon abundance is thus given by multiplying the yield by $m_\xi s_0/\rho_{\rm crit}$, giving
\begin{align}
\Omega_\xi h^2 =  m_\xi  \sum_i  \frac{4.6 \times 10^{27}}{g_{*s} \sqrt{g_*}}  \int_0^{T_R} \frac{{\cal C}_i (T)}{T^6} dT  \, .
\label{abundance1}
\end{align}

\paragraph{Collision terms} The form of the collision terms depend on the underlying production process. The relevant process here are flavor-violating decays $f_a \to f_b \xi$,  flavor-diagonal scatterings with photons (or gluons), $f_a \gamma \to f_a \xi$, fermion annihilations to CPons and photons (or gluons) $f_a \overline{f}_a \to \xi \gamma$, and finally scattering on Higgs bosons,  $f_a h \to f_a \xi$ and $f_a \overline{f}_a \to \xi h$ (in the high-energy limit). The respective collision terms read in terms of the  expressions in Section~\ref{production} (using Maxwell-Boltzmann instead of Fermi-Dirac distributions):
\begin{align}
{\cal C}_{f_a \to f_b \xi} & =  \frac{T m_a^2}{\pi^2} K_1 \left(\frac{m_a}{T} \right) \Gamma_{f_a \to f_b \xi} \, ,  \\
{\cal C}_{f_a \gamma \to f_a \xi} & =  \frac{T}{8 \pi^4}\int_{m_a^2}^\infty  \left(1- \frac{m_a^2}{s} \right)^2 s^{3/2}  \sigma_{f_a \gamma \to f_a \xi} (s) K_1 \left(\frac{\sqrt{s}}{T}\right) \, ds \, ,  \\
{\cal C}_{f_a \overline{f}_a \to \gamma \xi} & =\frac{T}{8 \pi^4 } \int_{4 m_a^2}^\infty  \left( 1 - \frac{4 m_a^2}{s} \right) s^{3/2} \sigma_{f_a \overline{f}_a \to \gamma \xi} (s) K_1 \left(\frac{\sqrt{s}}{T}\right) \, ds \, ,  \\
{\cal C}_{f_a \overline{f}_a \to h \xi} & = 2 {\cal C}_{f_a h  \to f_a \xi}  = \frac{T}{8 \pi^4 } \int_{0}^\infty   s^{3/2} \sigma^0_{f_a \overline{f}_a \to h \xi} K_1 \left(\frac{\sqrt{s}}{T}\right) \, ds \, ,
\end{align}
where the factor of 2 in the last line is due to the different spin degrees of freedom, and $K_1(x)$ denotes the modified Bessel function of the second kind. Expressions for gluons instead of photons are analogous. 
\paragraph{Relic abundancies} 
The temperature integral for the decays can readily be perfomed, and are dominated by the region where $T \approx m_a$, giving $\int K_1 (m_a/T)/T^5 dT \propto m_a^{-4}$ for dimensional reasons (as long as $T_R \gg m_a$). Similarly one can do the temperature integral for the  scatterings involving vector bosons, as the remaining $s$-integral is convergent in the UV due to  $\sigma_i \propto 1/s$ and can be done analytically. Instead the scattering on Higgs bosons is UV sensitive as the cross-section is constant, so one has first to perform the $s$-integral, giving $\int s^{3/2} K_1 (\sqrt{s}/T) \propto T^5$, and then perform the temperature integral which is linearly divergent and thus given by $T_{\max}$ (this justifies to work in the limit $\sqrt {s} \gg m_H, m_a$). The result for the integrated collision terms read 
\begin{align}
 \int_0^{\infty} \frac{{\cal C}_{f_a \to f_b \xi} (T)}{T^6} dT & = \frac{3}{2 \pi} \frac{\Gamma_{f_a \to f_b \xi}}{m_a^2} = \frac{3}{128 \pi^2 m_a \Lambda^2}  \left( |g_{ab}|^2 + |g_{ba}|^2 \right)\, , \\
 \int_0^{\infty} \frac{{\cal C}_{f_a \gamma \to f_a \xi} (T)}{T^6} dT & =  \frac{\alpha_{\rm em} Q_a^2}{168 \pi^3 m_a  \Lambda^2 }  \left( (63 \pi^2 - 600) y_{Saa}^2 + 16 y_{Paa}^2 \right) \, , \\
 \int_0^{\infty} \frac{{\cal C}_{f_a \overline{f}_a \to \gamma \xi} (T)}{T^6} dT & =   \frac{\alpha_{\rm em} Q_a^2}{160 \pi^2 m_a \Lambda^2 } \left( 13 y_{Saa}^2 + 15 y_{Paa}^2 \right) \, , \nonumber \\
  \int_0^{T_R} \frac{{\cal C}_{f_a \overline{f}_a \to h \xi} (T)}{T^6} dT & = \frac{4 T_R}{\pi^4} \sigma^0_{f_a \overline{f}_a \to h \xi} = \frac{T_R }{32 \pi^5 v^2 \Lambda^2 } |g_{aa}|^2 \, .
 \end{align}
 As the couplings scale as $y_{Paa} \sim y_{Saa} \sim g_{aa} \sim g_{ab} \sim g_{ba} \sim m_a$, the dominant contribution comes from the top quark (provided $T_R > m_t$), and the scattering on gluons instead of photons, which are analogous with $\alpha_{\rm em} Q^2 \to 24 \times \alpha_s (m_t)/6$ in $t  g \to t \xi$ and $\alpha_{\rm em} Q^2 \to 9 \times 4\alpha_s (m_t)/9$ in $t \overline{t} \to g \xi$, where the first factors compensate for the color averaging in the definition of the cross-section. The effective number of relativistic degrees of freedom in Eq.~\eqref{abundance1} is then given by $g_* = 106.75$ for all processes. Note that CPon  production from SUSY  scattering on gauge bosons is suppressed by at least $m_t/m_{\rm SUSY}$, but SUSY scattering on Higgs bosons is not, provided that $T_R > m_{\rm SUSY}$, i.e.  supersymmetric partners are in thermal equilibrium with the SM bath. Here we neglect this contribution, which depends on details of the supersymmetric spectrum, having in mind a scenario where this condition is not satisfied. 
 
Normalizing to the natural values of the couplings (see Section~\ref{QS}), we finally obtain  for the CPon relic abundancies, taking into account a possible factor of two for the charge-conjugated process, 
\begin{align}
\Omega_\xi h^2|_{t-{\rm decays}} & = 0.12 \left( \frac{m_\xi}{\keV} \right) \left( \frac{9.2 \times 10^9 \GeV}{\Lambda} \right)^2 \left( \frac{\sqrt{|g_{ct}|^2 + |g_{tc}|^2 }}{300 \GeV} \right)^2 \left( \frac{172 \GeV}{m_t} \right) \, ,\nonumber  \\
\Omega_\xi h^2|_{t-{\rm scat (IR)}} & = 0.12 \left( \frac{m_\xi}{\keV} \right) \left( \frac{1.4 \times 10^{11} \GeV}{\Lambda} \right)^2 \left( \frac{|g_{tt}|}{5.2 \TeV} \right)^2 \left( \frac{172 \GeV}{m_t} \right)  \left( \frac{\alpha_s (m_t)}{0.11} \right) \, , \nonumber \\
\Omega_\xi h^2|_{t-{\rm scat (UV)}} & = 0.12 \left( \frac{m_\xi}{\keV} \right) \left( \frac{1.4 \times 10^{11} \GeV}{\Lambda} \right)^2 \left( \frac{|g_{tt}|}{5.2 \TeV} \right)^2 \left( \frac{T_R}{3.1 \TeV} \right)  \, ,
\label{Oh2freezein}
\end{align}
where we restricted to $t \to c$ decays and took for simplicity $y_{Ptt} \approx y_{Stt} \approx |g_{tt}|/2$. From these results it is clear that decays are typically subleading to IR scattering, and UV scattering dominates over IR scattering if $T_R \gtrsim 3.1 \TeV$. These expressions are valid as long as $m_t \ll T_R < \Lambda$, while for reheating temperatures below the EW scale the main contribution to the freeze-in abundance occurs either from top scattering during the EMD epoch, which leads to a strong dilution of the resulting abundance (see e.g. Ref.~\cite{Silva-Malpartida:2024emu}), or from scattering of lighter fermions, for which the abundance is suppressed by small couplings. For simplicity we simply set the freeze-in contribution to zero for $T_R < 200 \GeV$, since it anyway has no impact in the relevant region of parameter space, see Section~\ref{results}. Moreover we need to restrict to $T_R < \Lambda$, as  the EFT we have used to calculate  CPon production rates is valid only for energies below the UV cutoff $\Lambda$.

\subsection{Limits on Warm Dark Matter}
\label{WDM}

While misalignment produces CPon DM with essentially vanishing momentum, CPons created by freeze-in have a large initial velocity and are initially free-streaming. This leads to a suppression of 
primordial fluctuations imprinted in the matter power spectrum at small scales, which can be constrained by looking at the spectra of distant quasars distorted by absorption in neutral hydrogen filaments that are assumed to trace the matter power spectrum, usually referred to as  the Lyman-$\alpha$ forest (Ly-$\alpha$)~\cite{Viel:2004bf, Boyarsky:2008xj}. This analysis yields a stringent lower bound on the warm DM mass $m_{\rm WDM}^{\rm min} \approx 5.3 \, \keV$~\cite{Viel:2013fqw,Baur:2015jsy,Irsic:2017ixq}, which can be relaxed to  $m_{\rm WDM}^{\rm min} \approx 3.5 \, \keV$ under more conservative assumptions. These Ly-$\alpha$ limits have been recasted for different freeze-in processes by computing the exact DM velocity distributionin Refs.~\cite{DEramo:2020gpr, Ballesteros:2020adh, Decant:2021mhj}. As we will show below, most relevant for our scenario is production via UV freeze-in which results in the ``Warm Dark Matter" (WDM) constraint~\cite{Ballesteros:2020adh}
\begin{align}
\label{WDMbound}
    m_\xi \gtrsim 7 \keV \left(\frac{m_{\rm WDM}^{\rm min}}{3 \,\text{keV}}\right)^{4/3}\left(\frac{106.75}{g_*(T_R)}\right)^{1/3} \, , 
\end{align}
where $m_{\rm WDM}\approx 3.5 \keV$ or $5.3 \keV$ for the conservative and stringent bounds, respectively, and $T_R$ denotes the reheating temperature. This bound gets relaxed if freeze-in  gives only a small fraction of a total abundance, which is mainly produced non-thermally via misalignment. While in this case one should re-asses the WDM bound in the given scenarios, here we refrain from this analysis (which is clearly beyond the scope of this work), and rather apply the lower limit in Eq.~\eqref{WDMbound}  only when the freeze-in fraction of the total relic abundance exceeds 1\%. Typically this happens only for  large reheating temperatures $T_R \gg \TeV$ (unless CPon couplings receive extremely large numerical enhancement factors with respect to the natural expectation), so that we can take $g_* (T_R) = 106.75$ and use for concreteness the WDM bound $m_\xi \ge 10 \keV$.

Finally we note that if the DM abundance is mainly generated via misalignment,   CPons with sufficiently small masses produced via decays and scattering of SM particles could still contribute to dark radiation, which is strongly constrained by BBN and CMB observations. Using these results of Refs.~\cite{DEramo:2024jhn, Badziak:2024qjg}, it is however clear that in our scenario these contributions are very efficiently suppressed by the UV scale  $\Lambda \gtrsim 10^{12} \GeV$, which makes the total amount of dark radiation negligible in the phenomenologically relevant regions of parameter space. 

\section{Concrete realizations}
\label{QS}
So far the framework we have considered is rather generic. To assess its capability of reproducing the DM abundance
while respecting all experimental bounds, we need to specialize our scenario. 
We consider two realizations, which provide an idea of the stability of our results against variations in the
underlying theory.
\subsection{Canonical K\"ahler potential}
\label{CK}
We start from the simple possibility of a theory with canonical K\"ahler potential. 
A general pattern of matrices of Yukawa couplings
that automatically verifies the condition in Eq.~\eqref{c2} is
\be
Y^A(\tau)=
\left(
\begin{array}{ccc}
0&0&c_{A13}\\
0&c_{A22}&c_{A23}(\tau)/x_A\\
c_{A31}&c_{A32}(\tau)/x_A&c_{A33}(\tau)/ x_A^2
\end{array}
\right)~~~~~~~~~(A=U,D,E)\,,
\ee
where $c_{A13}$, $c_{A22}$ and $c_{A31}$ are real constants and the $\tau$ dependence is carried by the lower-right triangular block. We have included a redundant real parameter $x_A$, one per each charge sector, that can be eliminated by redefining the 23, 32 and 33 entries. The advantage of this presentation is that when $x_A$ is smaller than one and all constants/functions $c_{Aab}$ are roughly of the same order of magnitude,
the singular values of $Y^A(\tau)$ are approximately given by $m_{A1}\approx c_A x_A^2$, $m_{A2}\approx c_A$ and $m_{A3}\approx c_A/x_A^2$, with 
a proportionality factor $c_A\approx |c_{Aab}|$ of about the same size. This is a good starting point to reproduce the observed hierarchy
in the charged fermion sector. The unitary matrices $U_{A^c}^T$ and $U_A$ that diagonalize $Y^A$ via $U_{A^c}^T Y^A U_A= \hat Y^A$ have the pattern:
\begin{align}
U_{A^c}^T\approx U_A\approx
\left(
\begin{array}{ccc}
1&x_A&x_A^2\\
x_A&1&x_A\\
x_A^2&x_A&1
\end{array}
\right)\,,
\end{align}
where only the order of magnitude of the entries has been displayed. The coefficients multiplying the off-diagonal terms
are indeed ratios of the quantities $c_{Aab}$, expected to be of order one if these quantities have approximately the same size. It follows that the generic pattern of the matrix of the CPon couplings, $\hat g_A=\langle\hat Y^A_\tau\rangle v_A$, evaluated in the basis where $\langle Y^A\rangle$ is diagonal, is
\begin{align}
\label{ansz}
\hat g_A
\approx
\frac{c_{A\tau}}{c_A}
\left(
\begin{array}{ccc}
m_{A1}&
\sqrt{m_{A1}m_{A2}}&
\sqrt{m_{A1}m_{A3}} \\
\sqrt{m_{A1}m_{A2}}&
m_{A2}&
\sqrt{m_{A2}m_{A3}}\\
\sqrt{m_{A1}m_{A3}}&
\sqrt{m_{A2}m_{A3}}&
m_{A3}
\end{array}
\right)
\,,
\end{align}
where we kept track of the fact that each entry is linearly proportional to a combination of the derivatives of the functions 
$c_{A23}(\tau)$, $c_{A32}(\tau)$ and $c_{A33}(\tau)$, that we have denoted by a common symbol $c_{A\tau}$. 
Our estimates depend on the ratio $c_{A\tau}/c_A$, describing the steepness of the functions  $c_{A23}(\tau)$, $c_{A32}(\tau)$ and $c_{A33}(\tau)$ evaluated at the minimum of the energy density. Without further knowledge of these
functions, we can adopt the  simple-minded ansatz
\be
\label{ansz2}
c_{A\tau}/c_A\approx 1\,.
\ee
Our assumptions are probably inadequate to reproduce, at a given energy scale, the precise values of the observed fermion masses and mixing angles. For instance, the $c_{Aab}$ input functions/parameters cannot be all exactly of the same size and they need some amount of tuning to match the experimental precision. 
Moreover, as we will see in a specific example, the near equality $c_{A\tau}/c_A\approx 1$ can be easily violated
by an order of magnitude. Keeping in mind all these caveats, here we will adopt Eqs.~\eqref{ansz} and (\ref{ansz2}) to
assess the allowed parameter space of the model. We also make use of the fermion masses evaluated at the scale of the $Z$ boson mass and listed in Table~\ref{fermass}. 
\begin{table}[h!]
\centering
\begin{tabular}{|c||c|} 
\hline
$m_d$~(GeV)&$(2.8\pm0.3)\times 10^{-3}$\\
\hline
$m_s$~(GeV)&$(54\pm3)\times 10^{-3}$\\
\hline
$m_b$~(GeV)&$2.85\pm0.03$\\
\hline
$m_u$~(GeV)&$(1.3\pm 0.5) \times 10^{-3}$\\
\hline
$m_c$~(GeV)&$0.63\pm0.02$\\
\hline
$m_t$~(GeV)&$171.7\pm1.6$\\
\hline
$m_e$~(GeV)&$(0.486654\pm0.000003)\times 10^{-3}$\\
\hline
$m_\mu$~(GeV)&$(10.2735\pm0.00003)\times 10^{-2}$\\
\hline
$m_\tau$~(GeV)&$1.74646\pm0.00002$\\
\hline
\end{tabular}
\caption{\label{fermass} 
Fermion masses renormalized at the scale $m_Z$, from Ref.~\cite{Antusch:2013jca}.}
\end{table}

\subsection{A modular-invariant model}
\label{MI}
As an alternative example, we consider the modular invariant model of Ref.~\cite{Feruglio:2023uof}, described in Appendix \ref{modular}. Each matter supermultiplet carries a weight $k_{Aa}$ and we consider the simple choice:
\begin{align}
k_{H_u}=k_{H_d}=0~~~~~~~~~~~~~~~~
k_{Q_a}=k_{U^c_a}=k_{D^c_a}=k_{L_a}=k_{E^c_a}=(-6,0,6),
\end{align}
which enforces the absence of gauge anomalies and guarantees a solution to the strong CP problem. 
Such a choice determines both the Yukawa couplings and the K\"ahler potential, in its minimal form.
The matrices of Yukawa couplings have entries that are modular forms of weight $k_{A^c_a}+k_{A_b}$:
\be\nn
Y^A(\tau)=\left(
\begin{array}{ccc}
0&0&c_{A13}\\
0&c_{A22}&c_{A23}E_6(\tau)\\
c_{A31}&c_{A32}E_6(\tau)&c_{A33}E_6(\tau)^2+{c'}_{A33}E_4(\tau)^3
\end{array}
\right)~~~~~~~~~~(A=U,D,E)\,,
\ee
where $c_{Aab}$ and $c'_{A33}$ are real constants, while $E_{4,6}(\tau)$ are the  Eisenstein modular forms of weight $4$ and $6$, see Appendix \ref{modular}.
The minimal K\"ahler metrics $\Omega^{c\dagger}_A \Omega^c_A$ and $\Omega^{\dagger}_A \Omega_A$ are determined by
\be
\Omega^c_A=y^{\frac{-k_{A^c_a}}{2}}\delta_{ab}~~~~~~~~~\Omega_A=y^\frac{-k_{A_a}}{2}\delta_{ab}~~~~~~~~~~~~~y=-i(\tau-\bar\tau)\,.
\ee
Plugging $Y^A(\tau)$ and $\Omega^c_A=\Omega_A$ into the general expressions of Eq.~\eqref{generalK}
we get the mass matrices $m_A$ and couplings $g_A$ of the model. We find
\be
m_{Aab}=\langle y^{k_{A^c_a}/2}Y^A_{ab}~ y^{k_{A_b}/2}\rangle\,.
\ee
We see that the role of the small parameter $x_A$ discussed in the previous section is played here by $1/y^3$, common
to all charged sectors. The couplings $g_A$ read~\cite{Ding:2020yen}:
\be
g_A=\frac{1}{y}\left[\langle y \times y^{k_{A^c_a}/2}Y^A_{\tau ab}~ y^{k_{A_b}/2}\rangle-i (k_{A^c a} m_{Aab}+m_{Aab} k_{A b})\right]\,.
\ee
Since $g_A$ only appears in the combination $g_A/\Lambda$ we will absorb the overall factor $1/y$ by redefining $\Lambda\to \langle y\rangle \Lambda$.
Choosing for convenience $\tau=1/8+i$, and fixing the input parameters $c_{Aab}$ and $c'_{A33}$ from a fit to
fermion masses and mixing angles (see Appendix \ref{modular}), we can compute the matrices $\hat g_A$ in the fermion mass basis. We get:
\begin{align}
\hat g_U=&~
\left(
\begin{array}{ccc}
 0.1189\, +0.085 i & 0 & -3.55+28.9 i \\
 0 & 0 & 0 \\
 -88.4+42.2 i & 0 & -15817-11387 i \\
\end{array}
\right) \GeV \nn\\
\hat g_D=&~
\left(
\begin{array}{ccc}
 0.168\, +0.531 i & 0.195\, -1.74 i & -4.59+36.3 i \\
 0.862\, +2.22 i & 0.366\, -7.48 i & -8.75+156 i \\
 -4.24-0.897 i & 11.1\, +7.93 i & -226-169 i \\
\end{array}
\right) \GeV \nn \\
\hat g_E=&~
\left(
\begin{array}{ccc}
 -0.00326+0.0232 i & 0 & 0.00157\, -0.0111 i \\
 0 & 0 & 0 \\
 -24.4+173 i & 0 & 11.7\, -83.1 i \\
\end{array}
\right) \GeV \, .
\label{fit}
\end{align}

\noindent
Scalar and pseudoscalar couplings are comparable and CP is violated in CPon interactions.
These couplings are dominated by the contribution proportional to $Y^A_{\tau ab}$, which is peculiar to this class of models. This is not the case for ALPs, whose
couplings to fermions are proportional to the PQ charges (the equivalent of $k_A$ and $k^c_A$).
Moreover, if we compare the couplings of this specific model with those estimated in Eqs.~\eqref{ansz} and (\ref{ansz2}), we see
that the ratio $c_{A\tau}/c_A$ can be larger than one, reaching values between one and two orders of magnitude
depending on the specific entries. 

\subsection{Numerical Results}
\label{results}

\begin{figure}[]
\centering
\includegraphics[width=0.49\linewidth]{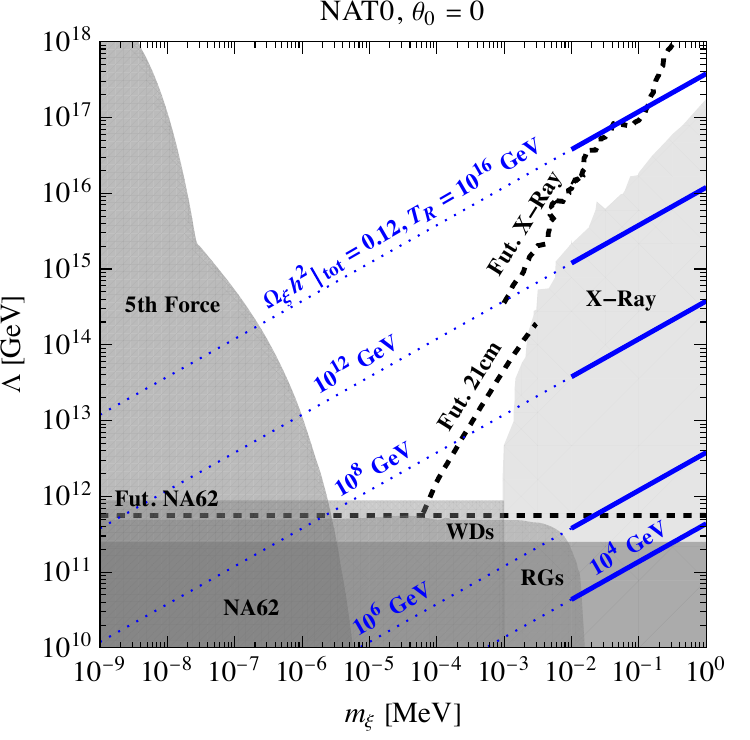}
\includegraphics[width=0.49\linewidth]{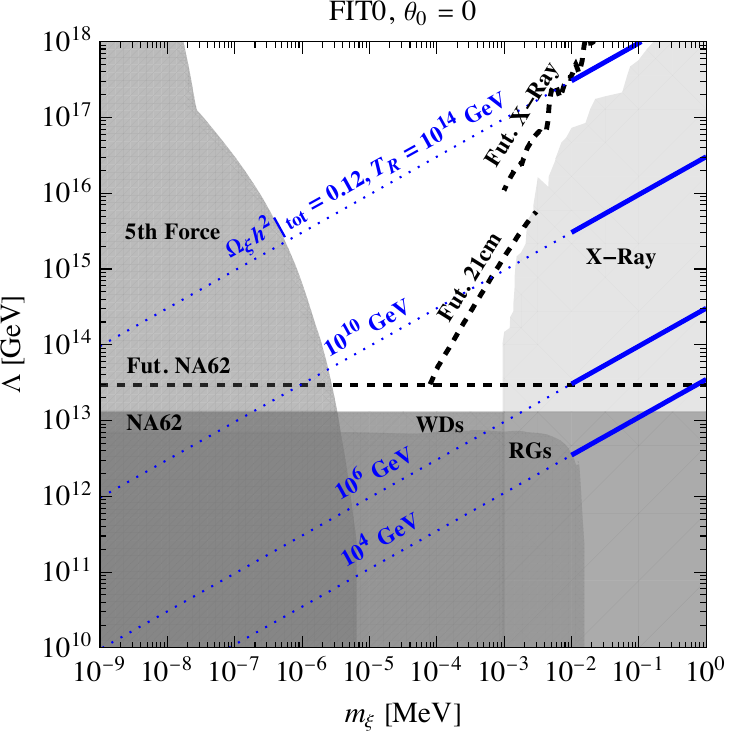}
\caption{Allowed parameter space for NAT0 (left panel) and FIT0 (right panel) scenarios, for vanishingly small values of the initial misalignment $\theta_0 = 0$.  As {\bf gray regions} we  show the constraints from Section 3, which are set by X-ray telescopes, inverse square law tests  ("5th Force") astrophysics ("RGs" and "WDs") and flavor experiments looking for $K \to \pi \xi$ decays ("NA62"). We also denote with {\bf black dashed contours} the expected sensitivity using the entire NA62 data set ("Fut. NA62"), the next generation of X-ray telescopes ("Fut. X-ray") and projected limits from the 21-cm power spectrum ("Fut. 21cm"). With {\bf blue contours} we indicated the regions of the parameter space where the observed DM abundance can be reproduced for the shown value of the reheating temperature $T_R$, with {\bf dotted blue lines} those excluded by the Lyman-$\alpha$ constraints on Warm DM, which require $m_\xi \gtrsim 10\keV$. 
\label{money1}}
\end{figure}

\begin{figure}[]
\centering
\includegraphics[width=0.49\linewidth]{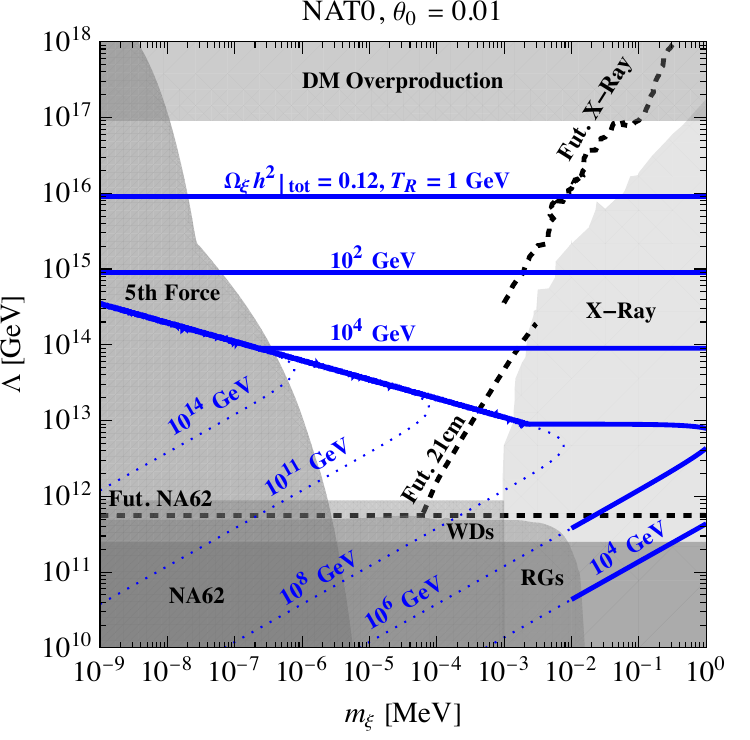}
\includegraphics[width=0.49\linewidth]{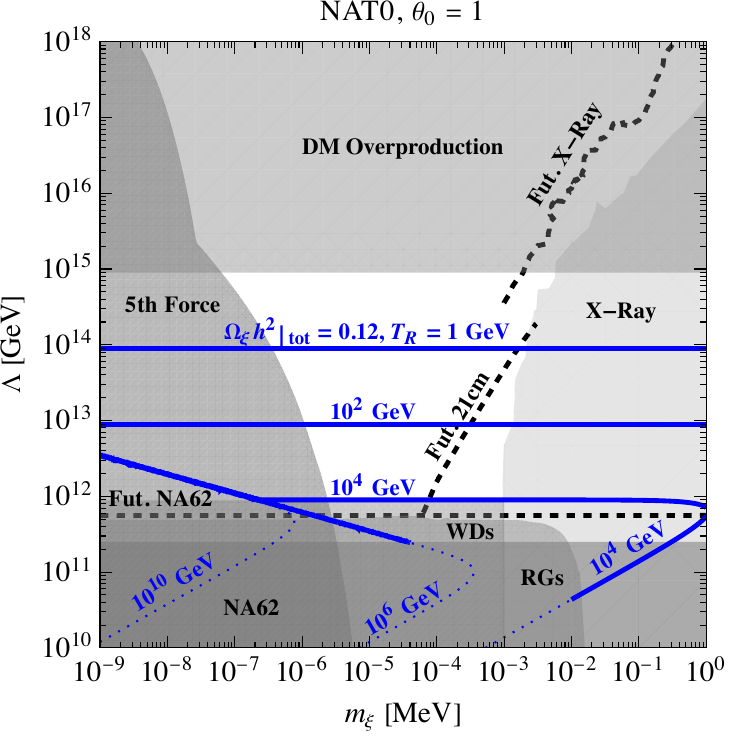}
\includegraphics[width=0.49\linewidth]{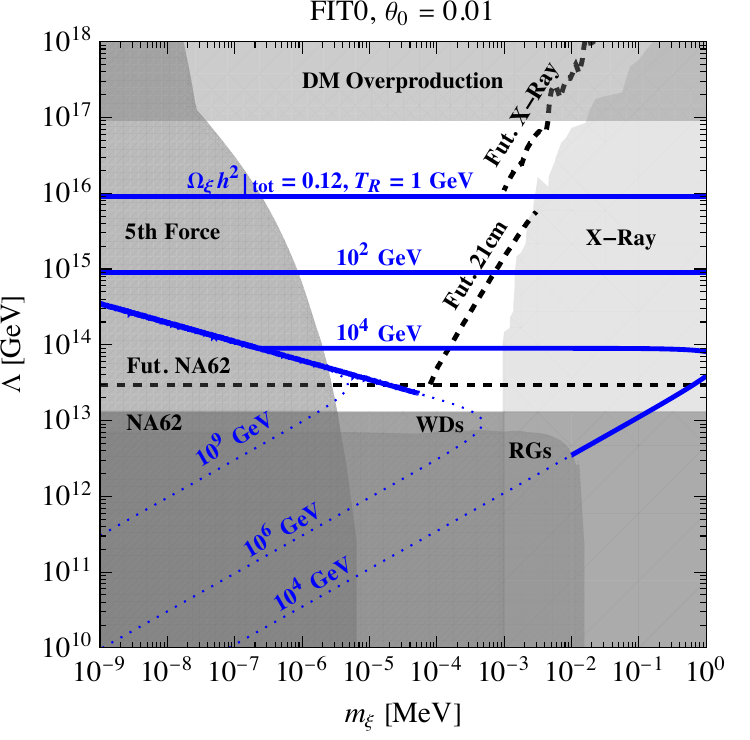}
\includegraphics[width=0.49\linewidth]{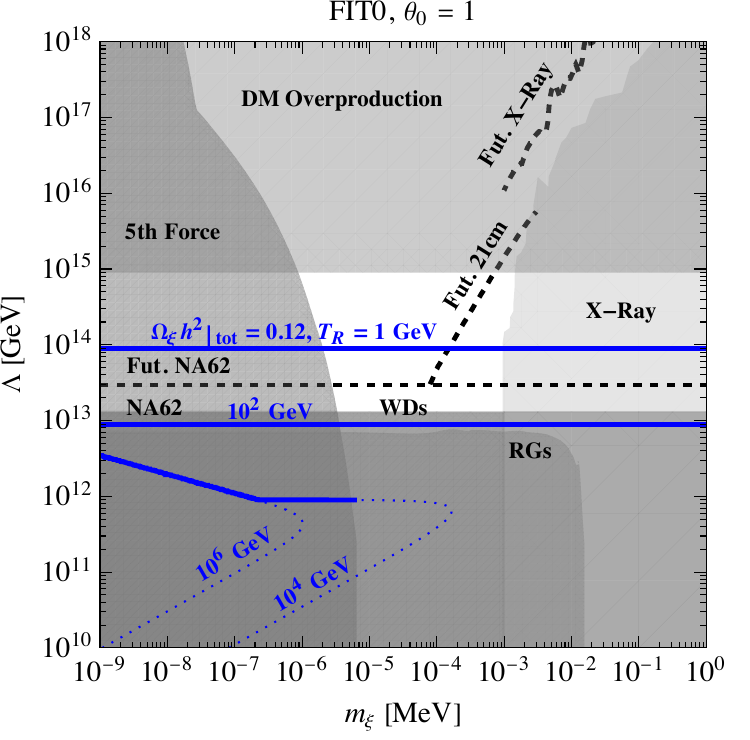}
\caption{Allowed parameter space for NAT0 (upper panel) and FIT0 (lower panel) scenarios, for indicated values of the initial misalignment $\theta_0 = \{0.01, 1\}$.  The {\bf gray regions} and {\bf black dashed contours} are as in Fig.~\ref{money1}. With {\bf blue contours} we indicated the regions of the parameter space where the observed DM abundance can be reproduced for the shown value of the reheating temperature $T_R$, with {\bf dotted blue lines}  by the Lyman-$\alpha$ constraints on Warm DM, which require $m_\xi \gtrsim 10\keV$ unless the dominant production is through misalignment (freeze-in contribution less than 1\%). The leading contribution to the abundance can be inferred with the scaling of the respective contour: $\Lambda \propto m_\xi^{1/2} T_R^{1/2}$ corresponds to freeze-in, 
$\Lambda \propto m_\xi^{-1/4} \theta_0^{-1}$ to RD misalignment and $\Lambda \propto T_R^{-1/2} \theta_0^{-1}$ (independent of $m_\xi$) to EMD misalignment.
\label{money2}}
\end{figure}

We now discuss the phenomenology of the CPon in the two explicit realizations detailed above, which we denote by ``NAT" (defined by Eq.~\eqref{ansz} with Eq.~\eqref{ansz2}) and ``FIT"  (defined by Eq.~\eqref{fit}). To specify the CPon couplings to fermions completely, we also have to fix the mixing angle $\alpha$ (cf.~Eq.~\eqref{ydef}). In the following we make the simple choice $\alpha =0$, other values change the couplings only marginally. We denote the  resulting scenarios by ``NAT0" and ``FIT0", and display the most relevant couplings in Table~\ref{Tab}. The  limits discussed in Section~\ref{pheno} from CPon decays ($y_{Pee},y_{See}$), flavor constraints ($y_{Sds},g_{e\mu}$), inverse square law tests  and star cooling ($y_{SNN},y_{See}$) then only depend on the CPon mass $m_\xi$ and the UV scale $\Lambda$, and are displayed  in the $m_\xi - \Lambda$ plane in Fig.~\ref{money1} and \ref{money2}, showing the excluded regions in gray. Note that in both scenarios  limits from HB star cooling and $\mu \to e \xi$ searches are sub-leading to $K \to \pi \xi$ constraints. The quantitative difference between the NAT0 and the FIT0 scenario can be easily understood from the numerical values in Table~\ref{Tab}, which are smaller in the NAT0 by a factor 10-100, which slightly relaxes the experimental constraints on the parameter space in the $m_\xi - \Lambda$ plane. 
\begin{table}[t]
\centering
\begin{tabular}{|c||cc|c|c|ccc|}
\hline 
 & $|y_{See}|$ & $|y_{Pee}|$ & $|y_{Sds}|$  & $|y_{SNN}| $  & $|g_{tt}|$ & $g_{ct,tc}$ & $g_{ut,tu}$ \\
\hline 
NAT0  & $3.4 \times 10^{-4}$ & $3.4 \times 10^{-4}$  & $0.027 $ &   $0.15$ & $2.4 \times 10^{2}$  & $21$  & $0.94$  \\
\hline
FIT0 &  $2.3 \times 10^{-3}$ & $1.6 \times 10^{-2}$  & $0.29$ &  $8.4$ & $1.9 \times 10^{4}$  & 0& $100$  \\
\hline
\end{tabular}
\caption{Values of phenomenologically relevant couplings in the explicit scenarios in units of GeV. We use the shorthand notation $g_{ij,ji} \equiv \sqrt{|g_{ij}|^2+|g_{ji}|^2}$.
\label{Tab}}
\end{table}

The CPon relic abundance instead depends on additional parameters. As discussed in Section~\ref{DM}, we are considering three independent contributions to the DM relic abundance: misalignment, IR-dominated freeze-in and UV-dominated freeze-in. For given couplings $\hat{g}_A$, the IR-freeze-in contribution depends only on the CPon mass $m_\xi$ and the UV scale  $\Lambda$, while the UV freeze-in contribution depends also on the reheating temperature $T_R$ (see Eq.~\eqref{Oh2freezein}). The misalignment abundance is independent of  fermion couplings, and besides $m_\xi$ and $\Lambda$  is set by the initial displacement $\theta_0$ and  $T_R$ (see Eq.~\eqref{Oh2misRD} and Eq.~\eqref{Oh2misEMD})~\footnote{The dependence on $\Lambda$ and $\theta_0$ is through the combination
$\xi_0=\Lambda\theta_0$.}.
Thus the total abundance depends on four parameters for a given realization (NAT0 or FIT0 as defined above). In each scenario and fixed $\theta_0$, we can determine the required value of $T_R$ needed to reproduce the observed DM abundance for given $m_\xi$ and $\Lambda$. This value follows from the total contributions to the relic abundance, which in the relevant scenarios are given by the following approximate expressions, where we restrict to the dominant contributions:
\begin{align}
\Omega_\xi h^2 |_{\rm tot}^{\rm NAT0} & = \left( \frac{m_\xi}{\keV} \right) \left( \frac{10^{12} \GeV}{\Lambda}  \right)^2 \left( 5.6 \times 10^{-6} + 0.17  \frac{T_R}{10^8 \GeV} \right) + \Omega_\xi h^2 |_{\rm mis} \, , \\
\Omega_\xi h^2 |_{\rm tot}^{\rm FIT0} & = \left( \frac{m_\xi}{\keV} \right) \left( \frac{10^{12} \GeV}{\Lambda}  \right)^2 \left( 3.6 \times 10^{-2} + 0.11  \frac{T_R}{10^4 \GeV} \right) + \Omega_\xi h^2 |_{\rm mis} \, ,
\end{align}
where the misalignment contribution is scenario-independent and given by~\footnote{In case of $T_R < 200 \GeV$ we only take into account the misalignment contribution, cf. Section~\ref{FI}.}
\begin{align}
\Omega_\xi h^2 |_{\rm mis}  & = 
0.12 \, \theta_0^2 \begin{cases} 
\left(\frac{\Lambda }{1.1\times 10^{11}\GeV}\right)^2   
\left( \frac{m_\xi}{\keV} \right)^{1/2} & T_R \ge 6.7 \times 10^5 \GeV \sqrt{m_\xi/\keV} \\
\left(\frac{\Lambda }{6.7 \times 10^{11}\GeV}\right)^2  \left(\frac{T_R}{18 \TeV}\right) & T_R < 6.7 \times 10^5 \GeV \sqrt{m_\xi/\keV}
\end{cases} \, .
\label{mis}
\end{align}
In Fig.~\ref{money1} and \ref{money2} we show in blue the contours where the relic abundance can be reproduced for the shown values of $T_R$, for three representative values of $\theta_0$ and the two scenarios. Dotted contours denote the regions excluded by the Warm DM bound below Eq.~\eqref{WDMbound}. In the following we discuss the qualitative feature of these contours.  


We start with the simplest scenario with vanishing (or very small) initial field values,  $\theta_0 = 0$ (Fig.~\ref{money1}). The CPon field sits near the minimum of the energy density, giving a limiting case that we discuss to isolate the feature of pure freeze-in. All DM contour lines follow the scaling  $\Lambda \propto m_\xi^{1/2}$. CPon DM is always produced with a large initial velocity and thus is constrained by structure formation (cf.~Section \ref{WDM}), excluding CPon masses below roughly 10 keV.  In order to generate the observed abundance one needs large values of $m_\xi$ and/or small values $\Lambda$, most of  which are in fact already excluded by a combination of flavor, astrophysical and X-ray constraints (see lower right contour) in both scenarios. For larger values of $\Lambda$ the correct abundance can be obtained by increasing $T_R$, and the parameter space is viable for values roughly above $T_R \approx 10^{12} \GeV$. In this region $\Lambda \propto m_\xi^{1/2} T_R^{1/2}$, so that all parameter space not excluded by other experiments allows to reproduce the observed abundance via UV freeze-in for sufficiently large $T_R < \Lambda$. As can be seen In Fig.~\ref{money1}, the associated parameter space with $T_R > \Lambda$ is excluded by Lyman-$\alpha$ constraints, so we do not explicitly impose this upper bound.

On the other hand we can consider $\theta_0 \approx 1$ (right panel of Fig.~\ref{money2}), where the dominant contribution in the relevant parameter space comes from misalignment, unless for very low values of $\Lambda$, which are in fact already excluded by WD and RG cooling and/or flavor constraints. For sufficiently low values of $T_R \lesssim \TeV$  the  main contribution to the CPon abundance is due to EMD misalignment. This contribution is then $m_\xi$-independent, and correspond to the horizontal  contour lines in Fig.~\ref{money2}, following $\Lambda \propto T_R^{-1/2} \theta_0^{-1}$. Because of the lower bound on $T_R$ of around 10 MeV from BBN,  there is a model-independent upper limit of $\Lambda \lesssim 10^{15} \GeV/ \theta_0$ above which the scenario is excluded by DM overproduction.  Larger reheating temperatures thus require lower values of $\Lambda$, and temperatures above $T_R \sim 10^4 \GeV$ ($\sim 10^2 \GeV$) are excluded for the NAT0 (FIT0) scenario. For reheating temperatures in the allowed window the DM abundance can be reproduced, and there is no WDM bound because DM is sufficiently cold. Although already excluded,  it is instructive to consider values of $T_R$ above  $10^4 \GeV$, where the relic abundance can be dominated by UV freeze-in for large values of $m_\xi$ (and low $\Lambda$), by EDM misalignment for intermediate values of $m_\xi$, and by RD misalignment for low values of $m_\xi$. In the latter region the contour line with the observed relic abundance scale as $\Lambda \propto  m_\xi^{-1/4}$, is $T_R-$independent and not subject to WDM constraints. However for sufficiently large $T_R$ the UV freeze-in contribution starts to become relevant, providing a second solution to the DM abundance with low $\Lambda$. While for the chosen value of $\theta_0 = 1$ the RD misalignment contribution  is entirely excluded by 5th force experiments, star cooling and/or flavor constraints, this contribution scales as  $\Lambda \propto \theta_0^{-1}$, so this case can  be made viable for slightly smaller values of $\theta_0$. 

For $\theta_0 = 0.01$ (left panel of Fig.~\ref{money2}) indeed  RD misalignment can give the dominant contribution to the observed abundance, without being in conflict with experimental constraints. This is particularly interesting, as in this case the abundance is insensitive to the precise value of $T_R$ in a broad range ($10^6 \GeV \lesssim T_R \lesssim 10^{14} \GeV$ for NAT0, and $10^6 \GeV \lesssim T_R \lesssim 10^{9} \GeV$ for FIT0). Also regions in the parameter space with dominant EMD misalignment are viable, for values of $\Lambda$ that are larger by factor 100 with respect to the $\theta = 1$, and an excluded region for DM overproduction that shrinks accordingly. Note that regions with dominant UV freeze-in are restricted to  $m_\xi \gtrsim 10 \keV$, and in fact excluded by X-ray constraints on decaying DM. As in Fig.~\ref{money1} we do not explicitly impose $T_R < \Lambda$ for the freeze-in contribution, as the associated parameter space with $T_R > \Lambda$ in the lower left part of the NAT0 scenario is anyway excluded by Lyman-$\alpha$ constraints.

\section{Conclusion}
\label{CL}

Solutions to the strong CP problem typically require extensions of the Standard Model that include
additional spin-zero particles in the spectrum. 
In the axion solution, the $\bar\theta$ parameter is promoted to a pseudoscalar field 
whose VEV is relaxed to zero by QCD dynamics. This mechanism works independently of the sources of
CP violation in the electroweak sector, whose nature remains
unexplored. In a wide range of parameter space, the axion is a viable candidate for cold DM,
currently under intense experimental search.

A different class of solutions assumes that the theory is invariant under CP, spontaneously broken
to deliver the observed CKM phase without affecting $\bar\theta$.
Such a breaking is achieved by the VEV of a (set of) complex spin-zero field(s),
upon which the Yukawa couplings of the theory depend.
We have considered this mechanism in the framework of a supersymmetric theory, where the field content
of the MSSM is minimally extended to include an extra gauge-singlet chiral supermultiplet.
Though supersymmetry is not a mandatory choice, it helps in accommodating
the relevant pattern of field-dependent Yukawa matrices, characterized by a constant determinant.  
The extra supermultiplet has nonrenormalizable interactions with the matter fields of the theory,
specified by a scale $\Lambda$. Without additional information about the dynamics of the
new supermultiplet, we cannot make any precise statement about the low-energy properties
of the theory. If the supermultiplet is very heavy, we have little hope of testing this scenario.

There are examples from string theory compactifications
where CP is a symmetry of the four-dimensional effective theory and the Yukawa couplings are dynamical quantities
depending on a set of moduli fields, some of which can be very light. 
Here we adopt the working assumption, which we are not able to justify in our bottom-up approach,
that one of the two spin-zero components of the extra supermultiplet is light,
with a mass $m_\xi$ that we treat as a free parameter.
We investigate under which circumstances such a light degree of freedom, the CPon, can act as Dark Matter.  

The CPon has CP-violating couplings to ordinary fermions, which are suppressed by the scale $\Lambda$
and entirely determined by the dynamical Yukawa couplings and by the K\"ahler potential of the theory.
Moreover, these couplings satisfy a sum rule, valid for both canonical 
and non-canonical K\"ahler potentials in a large class of scenarios. 
This sum rule implies that  CP-violating and charged lepton contributions to the photon decay rate are suppressed, and the decay amplitude into photons is dominated by the CP-conserving contribution from heavy quarks and light pseudoscalar mesons. The stringent bounds from X-ray telescopes on the CPon lifetime of order  $\tau_{\gamma \gamma} \lesssim 10^{29} {\rm \, sec}$, can be evaded if its mass is below 1 MeV (requiring $\Lambda$ below the Planck scale).
The scale $\Lambda$ is bounded from below (typically to be larger than $10^{12} \GeV$) by limits on the energy loss in White Dwarfs and Red Giants 
and by constraints on flavour-violating decays with CPon emission, most notably $K^+\to\pi^+ \xi$.
Importantly, the CPon mass cannot be arbitrarily small without affecting too much the inverse square law of gravity, with a typical lower bound of ${\cal O}$(meV).

For CPon production in the early universe, we have considered both misalignment and freeze-in, with a possible period of early matter domination preceeding the radiated dominated universe at a temperature $T_R$. In the allowed parameter space,  the CPon can easily saturate the observed Dark Matter abundance for suitable values of $T_R$, depending on the chosen values of CPon couplings and the inital misalignment. 

For a quantitative discussion, we have evaluated the relevant CPon couplings in two 
representative cases. First, we have provided an order-of-magnitude estimate, based on a typical pattern of Yukawa 
couplings delivering $\bar\theta=0$. 
Second, we have computed the CPon coupling constants in a model where the desired Yukawa matrices
are ensured by modular invariance. In this case, once the value of the CP-violating VEV
has been fixed, all free parameters can be derived from a fit to fermion masses and mixing angles,
with little residual uncertainty. 

We have analyzed the Dark Matter abundance in both scenarios, see Fig.~\ref{money1} and \ref{money2}.  Freeze-in is always the dominant mechanism when the initial value of the CPon field is very small. In this case, infrared freeze-in alone falls short to generate the observed relic abundance due to lower bounds on $\Lambda$, so that a large reheating temperature is needed to have a sufficiently large contribution from UV freeze-in. Given the present limits on warm Dark Matter and X-ray photons from DM decays, only a little portion of the parameter space is still available, see Fig.~\ref{money1}, partially in the reach of future X-ray missions.
Misalignment is the dominant mechanism as soon as the initial field value of the CPon is sufficiently large. For field values of the order of the UV scale, the lower bound on $\Lambda$ implies an upper bound on the reheating temperature of order $10^{2 \div 4} \GeV$, needed for a sufficient dilution of the CPon density in the EMD scenario. For lower initial field values instead the region of viable reheating temperatures opens up, allowing also regions in the parameter space where the dominant contribution to the DM abundance comes from usual RD misalignment.The abundance is then independent of the reheating temperature in a wide range of reheating temperatures between $10^4 \GeV \lesssim T_R \lesssim 10^{9 \div 14} \GeV$, for values of $\Lambda$ around $10^{13 \div 14} \GeV$, for the chosen value of the initial misalignment of $0.01 \Lambda$.

Though our scenario relies on a strong assumption, the lightness of the CPon, it has the attractive feature of linking three mysteries of fundamental interactions: the flavour puzzle, the origin of CP violation, and the nature of Dark Matter. Moreover, the presently allowed parameter space is
compact and will be partially explored by a variety of experimental probes in the near future, such as upcoming X-ray missions, 21cm cosmology, inverse square law tests of gravity, and laboratory searches for rare kaon decays.


\section*{Acknowledgements}
We thank Joerg Jaeckel, Luca Di Luzio, Hans Peter Nilles, Paolo Panci, Fernando Quevedo, Nicole Righi, Thomas Schwetz and Alessandro Strumia for useful discussions and comments on manuscript. We thank Gabriele Levati for his precious help in the computation of the CPon decay width.
The research of F.~F.~was supported in part by the INFN. F.~F.~thanks the KIT in Karlsruhe for hospitality in February 2024, when this project started. This work has received support from the European
Union's Horizon 2020 research and innovation programme under the Marie Sk{\l}odowska-Curie grant
agreement No 860881-HIDDeN and is partially supported by project B3a and C3b of the DFG-funded
Collaborative Research Center TRR257 ``Particle Physics Phenomenology after the Higgs Discovery".

\begin{appendices}
\section{CPon from Modular Invariance}
\label{modular}
We focus on a supersymmetric modular-invariant~\cite{Ferrara:1989bc,Ferrara:1989qb,Feruglio:2017spp} and CP-invariant~\cite{Baur:2019kwi,Novichkov:2019sqv,Baur:2019iai} theory. The Lagrangian $\mathscr{L}$ depends on a set of chiral supermultiplets $\phi$ comprising one (dimensionless) modulus $\tau$ (${\tt Im} \tau>0$) and the matter superfields $\phi_i$ of the MSSM. In a compact notation, $\mathscr{L}$ reads:
\be\label{susylag}
\mathscr{L}=\int d^2\theta d^2\bar\theta~ K(e^{2V} \varphi,\bar\varphi) + [\int d^2\theta~ w(\varphi) + h.c.] +[\frac{1}{16}\int d^2\theta~ f(\varphi)  W^a W^a+h.c.],
\ee 
where $\varphi$ collectively denotes all chiral supermultiplets. 
The K\"ahler potential $K$ is a real gauge-invariant function. The superpotential $w(\varphi)$ and the gauge kinetic functions $f_\alpha(\varphi)$ $(\alpha=1,2,3)$ are gauge-invariant analytic functions. The real and imaginary part of $f_3(\varphi)$
define the strong gauge coupling and the QCD angle~\footnote{We normalize the field strength as $W^a=2 W_B^a$, where 
$W_B^a$ is according to the definition in Ref.~\cite{Wess:1992cp}. It follows that $\int d^2\theta W^a W^a=-2 F^a_{\mu\nu}F^{a\mu\nu}+2 i
F^a_{\mu\nu} {\tilde F}^{a\mu\nu}$, where ${\tilde F}^{a\mu\nu}=1/2\epsilon^{\mu\nu\rho\sigma}F^a_{\rho\sigma}$.
From Eqs. (\ref{susylag}) and (\ref{kinef}) we get $1/16 \int d^2\theta~ f(\varphi)  W^a W^a+h.c.=-1/4g_s^2 F^a_{\mu\nu}F^{a\mu\nu}+\theta/32\pi^2 F^a_{\mu\nu} {\tilde F}^{a\mu\nu}$.
}:
\be\label{kinef}
f_3=\frac{1}{g_S^2}-i\frac{\theta_{\rm QCD}}{8\pi^2}\,.
\ee
Modular invariance means that $\mathscr{L}$ remains unchanged under $SL(2,\mathbbm{Z})$ 
transformations: 
\begin{align}
\tau\xrightarrow{\gamma}\gamma\tau=\frac{a \tau+b}{c\tau+d}~~~~~~~~~~~~
\phi_i\xrightarrow{\gamma} (c\tau+d)^{-k_i} \phi_i~~~~~~~~~~~~~~~~~
V\xrightarrow{\gamma}V,
\end{align}
where $a$, $b$, $c$, $d$ are integers obeying $ad-bc=1$ and $k_i$, the weights, are integer numbers. Up to modular transformations, under CP the multiplets transform as
\be
\tau\xrightarrow{CP}-\bar\tau~~~~~~~~~~~~\phi_i\xrightarrow{CP} \bar\phi_i~~~~~~~~~~~~V\xrightarrow{CP} \bar V\,.
\ee
The matter superfields $\phi_i$ are assumed to have either vanishing or negligible VEVs, compared to the VEV
of the modulus $\tau$. Since the theory is CP-invariant, the only source of CP violation is the VEV of the modulus $\tau$. Its CP conserving values are along the imaginary $\tau$ axis and along the border of the fundamental region $\Re(\tau)|\le 1/2$, $|\tau|\ge 1$.
\vskip 3.mm
\noindent
Consistently with modular and CP invariance, we can choose (real positive) constant gauge kinetic functions $f_\alpha(\varphi)=f_{0\alpha}$, such that the QCD angle $\theta_{\rm QCD}$ in Eq.~\eqref{kinef}  is zero.
We adopt a minimal K$\ddot{\mathrm{a}}$hler potential:
\begin{equation}
\label{eq:Kal}K(\phi,\bar{\phi}) = - \Lambda^2\log y+ \sum_{i}y ^{-k_i}\overline{\phi}_i\phi^i~~~~~~~~~~~y=-i(\tau-\overline{\tau})\,,
\end{equation}
where $\Lambda$ is the scale controlling the nonrenormalizable interaction of the modulus.
Working in the limit of massless neutrinos, the relevant part of the superpotential $w$ reads
\be
w(\varphi)=U^c_a H_u~ Y^U_{ab}(\tau)~ Q_b+D^c_a H_d ~Y^D_{ab}(\tau)~ Q_b+E^c_a H_d ~Y^E_{ab}(\tau)~ L_b+\mu(\tau) H_u H_d.
\ee
Modular invariance requires that the functions $Y^{U,D,E}_{ab}(\tau)$ and $\mu(\tau)$, assumed to be nonsingular, are
modular forms of weight $(k_{U^c_a}+k_{Q_b}+k_{H_u})$, $(k_{D^c_a}+k_{Q_b}+k_{H_d})$, $(k_{E^c_a}+k_{L_b}+k_{H_d})$
and $(k_{H_u}+k_{H_d})$, respectively.
We choose a basis of modular forms satisfying $Y_{ab}^A(-\bar\tau)=\overline{Y_{ab}^A(\tau)}$ and $\mu(-\bar\tau)=\overline{\mu(\tau)}$, so that any free parameter in the $w(\varphi)$ is constrained to be real.
\vskip 3.mm
\noindent
In the limit of exact supersymmetry, the physical angle $\bar\theta$, invariant under colored fermion chiral rotations, is 
\be\nn
\bar\theta=\theta_{\rm QCD}+\arg\det \mathscr{M}_U \mathscr{M}_D=\arg\det Y^U(\tau) ~Y^D(\tau)\,,
\ee
where $\mathscr{M}_{U,D}$ are the quark mass matrices and we
made use of $\theta_{\rm QCD}=0$ and the equality $\arg\det \mathscr{M}_U \mathscr{M}_D=\arg\det Y^U(\tau) ~Y^D(\tau)$,
which follows from the fact that the K\"ahler potential does not
affect the phase of the determinant of the mass matrices~\cite{Hiller:2001qg} and the VEVs of Higgs multiplets can be chosen real and positive.
The determinant $\det Y^U(\tau) ~Y^D(\tau)$ is a modular form of weight 
\begin{align}
\label{dettr}
k_{\rm det}=\sum_a(k_{U^c_a}+k_{D^c_a}+2k_{Q_a})+3 (k_{H_u}+k_{H_d}).
\end{align}
\noindent
Modular transformations act as local chiral rotations on canonically normalized fermion fields and
the weights of the matter multiplets should ensure the absence of mixed modular-gauge anomalies.
A simple solution to the set of conditions guaranteeing anomaly cancellation is~\cite{Feruglio:2023uof}
\begin{align}
\label{anosol}
k_{H_u}+k_{H_d}=0~~~~~~~~~~~~~~~~
k_{Q_a}=k_{U^c_a}=k_{D^c_a}=k_{L_a}=k_{E^c_a}=(-k,0,k),
\end{align}
\noindent
and we will adopt this choice, which implies $k_{\rm det}=0$.
The determinant $\det Y^u(\tau) ~Y^d(\tau)$ is a modular form of vanishing weight, which is a constant required to be real by CP invariance. If this constant is positive, which can be ensured by a proper choice of the free parameters, the physical angle $\bar\theta$ is zero. This result, valid in the limit of unbroken supersymmetry, gets corrected by a tolerable
amount if the mechanism of supersymmetry breaking does not introduce new phases and/or new flavour patterns~\cite{Hiller:2001qg,Hiller:2002um}.
The determinants of the up, down and electron mass matrices are also all real constants.
We remark that the condition for the absence of mixed modular-U(1)$_{QED}^2$, also satisfied by the solution
(\ref{anosol}), reads
\begin{align}
\label{mixU1em}
\sum_a \left[N_c Q_u^2 (k_{Q_a}+k_{U^c_a})+N_c Q_d^2 (k_{Q_a}+k_{D^c_a})+Q_e^2 (k_{L_a}+k_{E^c_a})\right]+(k_{H_u}+k_{H_d})=0.
\end{align}
Such a condition plays a role in the evaluation of the decay width of the modulus into two photons.
In this work, we further specify the choice in Eq.~\eqref{anosol}:
\begin{align}
k_{H_u}=k_{H_d}=0~~~~~~~~~~~~~~~~
k_{Q_a}=k_{U^c_a}=k_{D^c_a}=k_{L_a}=k_{E^c_a}=(-6,0,6)\,.
\end{align}
As a consequence, $\mu(\tau)=\mu$ is a real constant and the Yukawa couplings are given by
\be
Y^A(\tau)=\left(
\begin{array}{ccc}
0&0&c_{A13}\\
0&c_{A22}&c_{A23}E_6(\tau)\\
c_{A31}&c_{A32}E_6(\tau)&c_{A33}E_6(\tau)^2+{c'}_{A33}E_4(\tau)^3
\end{array}
\right)~~~~~~(A=U,D,E)\,,
\ee
where $c_{Aab}$ and $c'_{A33}$ are real constants, while $E_{4,6}(\tau)$ are the weight $(4,6)$ Eisenstein modular forms:
\begin{align}
E_4(\tau)=1+240 \sum _{n=1}^{\infty} \frac{n^3 \exp (2 \pi  i n \tau )}{1-\exp (2 \pi  i n \tau )}~~~~~~~
E_6(\tau)=1-504 \sum _{n=1}^{\infty} \frac{n^5 \exp (2 \pi  i n \tau )}{1-\exp (2 \pi  i n \tau )}\,.
\end{align}
From Eq.~\eqref{eq:Kal} we read the minimal K\"ahler metrics $\Omega^{c\dagger}_A \Omega^c_A$ and $\Omega^{\dagger}_A \Omega_A$, determined by
\be
\Omega^c_A=y^{\frac{-k_{A^c_a}}{2}}\delta_{ab}~~~~~~~~~\Omega_A=y^\frac{-k_{A_a}}{2}\delta_{ab}\,.
\ee
Plugging $Y^A(\tau)$ and $\Omega^c_A=\Omega_A$ into the general expressions of Eq.~\eqref{generalK}
we get the mass matrices $m_A$ and couplings $g_A$ of the model. We find
\be\label{starting}
m_A(\tau)=\left(
\begin{array}{ccc}
0&0&c_{A13}\\
0&c_{A22}&y^3 c_{A23}E_6(\tau)\\
c_{A31}&y^3 c_{A32}E_6(\tau)&y^6[c_{A33}E_6(\tau)^2+{c'}_{A33}E_4(\tau)^3]
\end{array}
\right)v_A\,,
\ee
where $v_U=v\sin\beta$ and $v_D=v_E=v\cos\beta$, $v\approx 174$ GeV and $\tau$ is evaluated at its VEV.
We see that the role of the small parameter $x_A$ discussed in Section \ref{QS} is played here by $1/y^3$, common
to all charged sectors. In this class of models ${\tt Im}(\tau)\ge \sqrt{3}/2$, and we have $1/y^3< 0.2$. The couplings $g_A$ read:
\be
g_{Aab}=\frac{1}{y}\left[\langle y \times y^{k_{A^c_a}/2}Y^A_{\tau ab}~ y^{k_{A_b}/2}\rangle-i (k_{A^c_a} m_{Aab}+m_{Aab} k_{A_b})\right]\,.
\ee
Since $g_A$ only appears in the combination $g_A/\Lambda$ we will absorb the overall factor $1/y$ by redefining $\Lambda\to \langle y\rangle \Lambda$.
There are more free parameters than observables. We reduce the number of free parameters
by choosing:
\be\nn
\tau=1/8+i\,.
\ee
Working in the vicinity of the imaginary unit has proved useful~\cite{Feruglio:2021dte,Novichkov:2021evw,Feruglio:2022koo,Feruglio:2023mii,Ding:2024xhz} in model building. Moreover,
we have simple approximate expressions for the quantities 
of interest by making an expansion in the small parameter $x=1/y^3=0.125$.
In addition, we choose:
\begin{align}
\nn
c_{U23}=c_{U32}=c_{E23}=c_{E32}=&~0~~~~~~~~~~~~~~~~~~~~~~~~~~~~~~\tan\beta=10\nn\\
\arg[c_{U33}E_6(\tau)^2+{c'}_{U33}E_4(\tau)^3]=&~\arg[c_{D33}E_6(\tau)^2+{c'}_{D33}E_4(\tau)^3]~~~~~~~{\rm at}~\tau=1/8+i\,.\nn
\end{align}
We are left with a total of 16 real parameters (vs. 13 observables), to be determined by maximizing the agreement with fermion masses, mixing angles and CKM phase. The experimental data and their errors, renormalized at the scale $m_Z$ and taken from Ref.~\cite{Antusch:2013jca}, are 
collected in Table \ref{compare}. 
\begin{table}[h!]
\centering
\begin{tabular}{|c|c|c|c|c|c|c|c|} 
\hline
$A$&$c_{A13}$&$c_{A22}$&$c_{A23}$&$c_{A31}$&$c_{A32}$&$c_{A33}$&${c'}_{A33}$\\
\hline
$D$&$1.063$&$0.541$&$-1.278$&$0.143$&$-0.109$&$-1.194$&$-0.395$\\
\hline
$U$&$0.148$&$0.362$&$0$&$0.498$&$0$&$-7.186$&$-2.379$\\
\hline
$E$&$0.00312$&$0.593$&$0$&$9.09$&$0$&$0.756$&$0.192$\\
\hline
\end{tabular}
\caption{\label{candcp} Coefficients $c_{Aab}$ and  ${c'}_{Aab}$ in units 0.01. We choose $v_U=173.3$ GeV
and $v_D=17.3$ GeV.
}
\end{table}
We show the best fit values of $c_{Aab}$,  ${c'}_{Aab}$ in Table \ref{candcp}.
In Table \ref{compare} we display the model predictions evaluated at the best-fit point and we compare them with the
experimental values.
\begin{table}[h!]
\centering
\begin{tabular}{|c||c|c|} 
\hline
&Experiment&Model\\
\hline
$m_d$~(GeV)&$(2.8\pm0.3)\times 10^{-3}$&$2.6\times 10^{-3}$\\
\hline
$m_s$~(GeV)&$(54\pm3)\times 10^{-2}$&$50\times 10^{-3}$\\
\hline
$m_b$~(GeV)&$2.85\pm0.03$&$3.19$\\
\hline
$m_u$~(GeV)&$(1.3\pm 0.5) \times 10^{-3}$&$1.3\times 10^{-3}$\\
\hline
$m_c$~(GeV)&$0.63\pm0.02$&$0.63$\\
\hline
$m_t$~(GeV)&$171.7\pm1.6$&$171.7$\\
\hline
$\sin\theta_{12}$&$0.22540\pm0.00070$&$0.213$\\
\hline
$\sin\theta_{23}$&$0.0420\pm0.0006$&$0.045$\\
\hline
$\sin\theta_{13}$&$0.00364\pm0.00013$&$0.0019$\\
\hline
$\sin\delta$&$0.93\pm0.02$&$0.88$\\
\hline
\hline
$m_e$~(GeV)&$(0.486654\pm0.000003)\times 10^{-3}$&$0.486653\times 10^{-3}$\\
\hline
$m_\mu$~(GeV)&$(10.2735\pm0.00003)\times 10^{-2}$&$10.2735\times 10^{-2}$\\
\hline
$m_\tau$~(GeV)&$1.74646\pm0.00002$&$1.74645$\\
\hline
\end{tabular}
\caption{\label{compare} 
Left column: fermion masses, mixing angles, and CKM phase, renormalized at the scale $m_Z$, from Ref.~\cite{Antusch:2013jca}. Right column: model predictions from $\tau=1/8+i$, $\tan\beta=10$ and the input parameters of Table \ref{candcp}.}
\end{table}
The agreement in the lepton sector is very good, while in the quark sector is decent, with some discrepancies due to the use of approximate theoretical expressions for the quantities of interest.
Evaluating the modulus coupling constants in the basis where the mass matrices are diagonal we get
we find
\begin{align}
\hat g_U=&~
\left(
\begin{array}{ccc}
 0.1189\, +0.085 i & 0 & -3.55+28.9 i \\
 0 & 0 & 0 \\
 -88.4+42.2 i & 0 & -15817-11387 i \\
\end{array}
\right)  \GeV
\nn\\
\hat g_D=&~
\left(
\begin{array}{ccc}
 0.168\, +0.531 i & 0.195\, -1.74 i & -4.59+36.3 i \\
 0.862\, +2.22 i & 0.366\, -7.48 i & -8.75+156 i \\
 -4.24-0.897 i & 11.1\, +7.93 i & -226-169 i \\
\end{array}
\right) \GeV \\
\hat g_E=&~
\left(
\begin{array}{ccc}
 -0.00326+0.0232 i & 0 & 0.00157\, -0.0111 i \\
 0 & 0 & 0 \\
 -24.4+173 i & 0 & 11.7\, -83.1 i \\
\end{array}
\right) \GeV \,.\nn
\end{align}

\end{appendices}
\newpage

\bibliographystyle{JHEP2}
\bibliography{draft2025}

\providecommand{\href}[2]{#2}\begingroup\raggedright\begin{thebibliography}{100}

\bibitem{Cirelli:2024ssz}
M.~Cirelli, A.~Strumia, and J.~Zupan, {\it {Dark Matter}},
  \href{http://arxiv.org/abs/2406.01705}{{\tt arXiv:2406.01705}}.

\bibitem{Weinberg:1977ma}
S.~Weinberg, {\it {A New Light Boson?}},  {\em Phys. Rev. Lett.} {\bf 40}
  (1978) 223--226.

\bibitem{Wilczek:1977pj}
F.~Wilczek, {\it {Problem of Strong $P$ and $T$ Invariance in the Presence of
  Instantons}},  {\em Phys. Rev. Lett.} {\bf 40} (1978) 279--282.

\bibitem{Peccei:1977hh}
R.~D. Peccei and H.~R. Quinn, {\it {CP Conservation in the Presence of
  Instantons}},  {\em Phys. Rev. Lett.} {\bf 38} (1977) 1440--1443.

\bibitem{Preskill:1982cy}
J.~Preskill, M.~B. Wise, and F.~Wilczek, {\it {Cosmology of the Invisible
  Axion}},  {\em Phys. Lett. B} {\bf 120} (1983) 127--132.

\bibitem{Abbott:1982af}
L.~F. Abbott and P.~Sikivie, {\it {A Cosmological Bound on the Invisible
  Axion}},  {\em Phys. Lett. B} {\bf 120} (1983) 133--136.

\bibitem{Dine:1982ah}
M.~Dine and W.~Fischler, {\it {The Not So Harmless Axion}},  {\em Phys. Lett.
  B} {\bf 120} (1983) 137--141.

\bibitem{DiLuzio:2020wdo}
L.~Di~Luzio, M.~Giannotti, E.~Nardi, and L.~Visinelli, {\it {The landscape of
  QCD axion models}},  {\em Phys. Rept.} {\bf 870} (2020) 1--117,
  [\href{http://arxiv.org/abs/2003.01100}{{\tt arXiv:2003.01100}}].

\bibitem{Kamionkowski:1992mf}
M.~Kamionkowski and J.~March-Russell, {\it {Planck scale physics and the
  Peccei-Quinn mechanism}},  {\em Phys. Lett. B} {\bf 282} (1992) 137--141,
  [\href{http://arxiv.org/abs/hep-th/9202003}{{\tt hep-th/9202003}}].

\bibitem{Davidson:1981zd}
A.~Davidson and K.~C. Wali, {\it {MINIMAL FLAVOR UNIFICATION VIA
  MULTIGENERATIONAL PECCEI-QUINN SYMMETRY}},  {\em Phys. Rev. Lett.} {\bf 48}
  (1982) 11.

\bibitem{Wilczek:1982rv}
F.~Wilczek, {\it {Axions and Family Symmetry Breaking}},  {\em Phys. Rev.
  Lett.} {\bf 49} (1982) 1549--1552.

\bibitem{Berezhiani:1989fp}
Z.~G. Berezhiani and M.~Y. Khlopov, {\it {Cosmology of Spontaneously Broken
  Gauge Family Symmetry}},  {\em Z. Phys. C} {\bf 49} (1991) 73--78.

\bibitem{Ema:2016ops}
Y.~Ema, K.~Hamaguchi, T.~Moroi, and K.~Nakayama, {\it {Flaxion: a minimal
  extension to solve puzzles in the standard model}},  {\em JHEP} {\bf 01}
  (2017) 096, [\href{http://arxiv.org/abs/1612.05492}{{\tt arXiv:1612.05492}}].

\bibitem{Calibbi:2016hwq}
L.~Calibbi, F.~Goertz, D.~Redigolo, R.~Ziegler, and J.~Zupan, {\it {Minimal
  axion model from flavor}},  {\em Phys. Rev. D} {\bf 95} (2017), no.~9 095009,
  [\href{http://arxiv.org/abs/1612.08040}{{\tt arXiv:1612.08040}}].

\bibitem{Feruglio:2023uof}
F.~Feruglio, A.~Strumia, and A.~Titov, {\it {Modular invariance and the QCD
  angle}},  {\em JHEP} {\bf 07} (2023) 027,
  [\href{http://arxiv.org/abs/2305.08908}{{\tt arXiv:2305.08908}}].

\bibitem{Feruglio:2024ytl}
F.~Feruglio, M.~Parriciatu, A.~Strumia, and A.~Titov, {\it {Solving the strong
  CP problem without axions}},  {\em JHEP} {\bf 08} (2024) 214,
  [\href{http://arxiv.org/abs/2406.01689}{{\tt arXiv:2406.01689}}].

\bibitem{Nelson:1983zb}
A.~E. Nelson, {\it {Naturally Weak CP Violation}},  {\em Phys. Lett. B} {\bf
  136} (1984) 387--391.

\bibitem{Barr:1984qx}
S.~M. Barr, {\it {Solving the Strong CP Problem Without the Peccei-Quinn
  Symmetry}},  {\em Phys. Rev. Lett.} {\bf 53} (1984) 329.

\bibitem{Georgi:1978xz}
H.~Georgi, {\it {A Model of Soft CP Violation}},  {\em Hadronic J.} {\bf 1}
  (1978) 155.

\bibitem{Hall:2024xbd}
L.~Hall, C.~A. Manzari, and B.~Noether, {\it {Strong CP and Flavor in
  Multi-Higgs Theories}},  \href{http://arxiv.org/abs/2407.14585}{{\tt
  arXiv:2407.14585}}.

\bibitem{Ferro-Hernandez:2024snl}
R.~Ferro-Hernandez, S.~Morisi, and E.~Peinado, {\it {Axionless strong CP
  problem solution: the spontaneous CP violation case}},
  \href{http://arxiv.org/abs/2407.18161}{{\tt arXiv:2407.18161}}.

\bibitem{Antusch:2013rla}
S.~Antusch, M.~Holthausen, M.~A. Schmidt, and M.~Spinrath, {\it {Solving the
  Strong CP Problem with Discrete Symmetries and the Right Unitarity
  Triangle}},  {\em Nucl. Phys. B} {\bf 877} (2013) 752--771,
  [\href{http://arxiv.org/abs/1307.0710}{{\tt arXiv:1307.0710}}].

\bibitem{Vecchi:2014hpa}
L.~Vecchi, {\it {Spontaneous CP violation and the strong CP problem}},  {\em
  JHEP} {\bf 04} (2017) 149, [\href{http://arxiv.org/abs/1412.3805}{{\tt
  arXiv:1412.3805}}].

\bibitem{Dine:2015jga}
M.~Dine and P.~Draper, {\it {Challenges for the Nelson-Barr Mechanism}},  {\em
  JHEP} {\bf 08} (2015) 132, [\href{http://arxiv.org/abs/1506.05433}{{\tt
  arXiv:1506.05433}}].

\bibitem{Schwichtenberg:2018aqc}
J.~Schwichtenberg, P.~Tremper, and R.~Ziegler, {\it {A grand-unified
  Nelson\textendash{}Barr model}},  {\em Eur. Phys. J. C} {\bf 78} (2018),
  no.~11 910, [\href{http://arxiv.org/abs/1802.08109}{{\tt arXiv:1802.08109}}].

\bibitem{Valenti:2021rdu}
A.~Valenti and L.~Vecchi, {\it {The CKM phase and $ \overline{\theta} $ in
  Nelson-Barr models}},  {\em JHEP} {\bf 07} (2021), no.~203 203,
  [\href{http://arxiv.org/abs/2105.09122}{{\tt arXiv:2105.09122}}].

\bibitem{Valenti:2021xjp}
A.~Valenti and L.~Vecchi, {\it {Super-soft CP violation}},  {\em JHEP} {\bf 07}
  (2021), no.~152 152, [\href{http://arxiv.org/abs/2106.09108}{{\tt
  arXiv:2106.09108}}].

\bibitem{Nakagawa:2024ddd}
S.~Nakagawa, Y.~Nakai, and Y.~Wang, {\it {Spontaneous CP violation in
  supersymmetric QCD}},  {\em JHEP} {\bf 09} (2024) 105,
  [\href{http://arxiv.org/abs/2406.01260}{{\tt arXiv:2406.01260}}].

\bibitem{Hall:2009bx}
L.~J. Hall, K.~Jedamzik, J.~March-Russell, and S.~M. West, {\it {Freeze-In
  Production of FIMP Dark Matter}},  {\em JHEP} {\bf 03} (2010) 080,
  [\href{http://arxiv.org/abs/0911.1120}{{\tt arXiv:0911.1120}}].

\bibitem{Panci:2022wlc}
P.~Panci, D.~Redigolo, T.~Schwetz, and R.~Ziegler, {\it {Axion dark matter from
  lepton flavor-violating decays}},  {\em Phys. Lett. B} {\bf 841} (2023)
  137919, [\href{http://arxiv.org/abs/2209.03371}{{\tt arXiv:2209.03371}}].

\bibitem{Aghaie:2024jkj}
M.~Aghaie, G.~Armando, A.~Conaci, et~al., {\it {Axion dark matter from heavy
  quarks}},  {\em Phys. Lett. B} {\bf 856} (2024) 138923,
  [\href{http://arxiv.org/abs/2404.12199}{{\tt arXiv:2404.12199}}].

\bibitem{Petcov:2024vph}
S.~T. Petcov and M.~Tanimoto, {\it {$A_4$ modular invariance and the strong CP
  problem}},  {\em Eur. Phys. J. C} {\bf 84} (2024), no.~9 914,
  [\href{http://arxiv.org/abs/2404.00858}{{\tt arXiv:2404.00858}}].

\bibitem{Penedo:2024gtb}
J.~T. Penedo and S.~T. Petcov, {\it {Finite modular symmetries and the strong
  CP problem}},  {\em JHEP} {\bf 10} (2024) 172,
  [\href{http://arxiv.org/abs/2404.08032}{{\tt arXiv:2404.08032}}].

\bibitem{Cvetic:1991qm}
M.~Cvetic, A.~Font, L.~E. Ibanez, D.~Lust, and F.~Quevedo, {\it {Target space
  duality, supersymmetry breaking and the stability of classical string
  vacua}},  {\em Nucl. Phys. B} {\bf 361} (1991) 194--232.

\bibitem{Ishiguro:2020nuf}
K.~Ishiguro, T.~Kobayashi, and H.~Otsuka, {\it {Spontaneous CP violation and
  symplectic modular symmetry in Calabi-Yau compactifications}},  {\em Nucl.
  Phys. B} {\bf 973} (2021) 115598,
  [\href{http://arxiv.org/abs/2010.10782}{{\tt arXiv:2010.10782}}].

\bibitem{Novichkov:2022wvg}
P.~P. Novichkov, J.~T. Penedo, and S.~T. Petcov, {\it {Modular flavour
  symmetries and modulus stabilisation}},  {\em JHEP} {\bf 03} (2022) 149,
  [\href{http://arxiv.org/abs/2201.02020}{{\tt arXiv:2201.02020}}].

\bibitem{Leedom:2022zdm}
J.~M. Leedom, N.~Righi, and A.~Westphal, {\it {Heterotic de Sitter beyond
  modular symmetry}},  {\em JHEP} {\bf 02} (2023) 209,
  [\href{http://arxiv.org/abs/2212.03876}{{\tt arXiv:2212.03876}}].

\bibitem{Higaki:2024pql}
T.~Higaki, T.~Kobayashi, K.~Nasu, and H.~Otsuka, {\it {Spontaneous CP violation
  and partially broken modular flavor symmetries}},  {\em JHEP} {\bf 09} (2024)
  024, [\href{http://arxiv.org/abs/2405.18813}{{\tt arXiv:2405.18813}}].

\bibitem{Banks:1996ss}
T.~Banks and M.~Dine, {\it {Couplings and scales in strongly coupled heterotic
  string theory}},  {\em Nucl. Phys. B} {\bf 479} (1996) 173--196,
  [\href{http://arxiv.org/abs/hep-th/9605136}{{\tt hep-th/9605136}}].

\bibitem{Choi:1998dw}
K.~Choi, E.~J. Chun, and H.~B. Kim, {\it {Cosmology of light moduli}},  {\em
  Phys. Rev. D} {\bf 58} (1998) 046003,
  [\href{http://arxiv.org/abs/hep-ph/9801280}{{\tt hep-ph/9801280}}].

\bibitem{Cicoli:2012sz}
M.~Cicoli, M.~Goodsell, and A.~Ringwald, {\it {The type IIB string axiverse and
  its low-energy phenomenology}},  {\em JHEP} {\bf 10} (2012) 146,
  [\href{http://arxiv.org/abs/1206.0819}{{\tt arXiv:1206.0819}}].

\bibitem{Cicoli:2021gss}
M.~Cicoli, V.~Guidetti, N.~Righi, and A.~Westphal, {\it {Fuzzy Dark Matter
  candidates from string theory}},  {\em JHEP} {\bf 05} (2022) 107,
  [\href{http://arxiv.org/abs/2110.02964}{{\tt arXiv:2110.02964}}].

\bibitem{Cicoli:2023opf}
M.~Cicoli, J.~P. Conlon, A.~Maharana, et~al., {\it {String cosmology: From the
  early universe to today}},  {\em Phys. Rept.} {\bf 1059} (2024) 1--155,
  [\href{http://arxiv.org/abs/2303.04819}{{\tt arXiv:2303.04819}}].

\bibitem{Funakoshi:2024yxg}
S.~Funakoshi, J.~Kawamura, T.~Kobayashi, K.~Nasu, and H.~Otsuka, {\it {Moduli
  stabilization and light axion by Siegel modular forms}},
  \href{http://arxiv.org/abs/2409.19261}{{\tt arXiv:2409.19261}}.

\bibitem{Higaki:2024jdk}
T.~Higaki, J.~Kawamura, and T.~Kobayashi, {\it {Finite modular axion and
  radiative moduli stabilization}},  {\em JHEP} {\bf 04} (2024) 147,
  [\href{http://arxiv.org/abs/2402.02071}{{\tt arXiv:2402.02071}}].

\bibitem{Baur:2024lcc}
A.~Baur, M.-C. Chen, V.~Knapp-Perez, and S.~Ramos-Sanchez, {\it {Modular
  flavored dark matter}},  {\em JHEP} {\bf 12} (2024) 091,
  [\href{http://arxiv.org/abs/2409.02178}{{\tt arXiv:2409.02178}}].

\bibitem{Chowdhury:2018tzw}
D.~Chowdhury, E.~Dudas, M.~Dutra, and Y.~Mambrini, {\it {Moduli Portal Dark
  Matter}},  {\em Phys. Rev. D} {\bf 99} (2019), no.~9 095028,
  [\href{http://arxiv.org/abs/1811.01947}{{\tt arXiv:1811.01947}}].

\bibitem{Dine:2024bxv}
M.~Dine, G.~Perez, W.~Ratzinger, and I.~Savoray, {\it {Nelson-Barr ultralight
  dark matter}},  {\em Phys. Rev. D} {\bf 111} (2025), no.~1 015049,
  [\href{http://arxiv.org/abs/2405.06744}{{\tt arXiv:2405.06744}}].

\bibitem{Grzadkowski:2018nbc}
B.~Grzadkowski and D.~Huang, {\it {Spontaneous $CP$-Violating Electroweak
  Baryogenesis and Dark Matter from a Complex Singlet Scalar}},  {\em JHEP}
  {\bf 08} (2018) 135, [\href{http://arxiv.org/abs/1807.06987}{{\tt
  arXiv:1807.06987}}].

\bibitem{Hiller:2001qg}
G.~Hiller and M.~Schmaltz, {\it {Solving the Strong CP Problem with
  Supersymmetry}},  {\em Phys. Lett. B} {\bf 514} (2001) 263--268,
  [\href{http://arxiv.org/abs/hep-ph/0105254}{{\tt hep-ph/0105254}}].

\bibitem{Hiller:2002um}
G.~Hiller and M.~Schmaltz, {\it {Strong Weak CP Hierarchy from
  Nonrenormalization Theorems}},  {\em Phys. Rev. D} {\bf 65} (2002) 096009,
  [\href{http://arxiv.org/abs/hep-ph/0201251}{{\tt hep-ph/0201251}}].

\bibitem{Ellis:1978hq}
J.~R. Ellis and M.~K. Gaillard, {\it {Strong and Weak CP Violation}},  {\em
  Nucl. Phys. B} {\bf 150} (1979) 141--162.

\bibitem{Khriplovich:1985jr}
I.~B. Khriplovich, {\it {Quark Electric Dipole Moment and Induced $\theta$ Term
  in the {Kobayashi-Maskawa} Model}},  {\em Phys. Lett. B} {\bf 173} (1986)
  193--196.

\bibitem{Kusenko:2012ch}
A.~Kusenko, M.~Loewenstein, and T.~T. Yanagida, {\it {Moduli dark matter and
  the search for its decay line using Suzaku X-ray telescope}},  {\em Phys.
  Rev. D} {\bf 87} (2013), no.~4 043508,
  [\href{http://arxiv.org/abs/1209.6403}{{\tt arXiv:1209.6403}}].

\bibitem{Quevedo:2014xia}
F.~Quevedo, {\it {Local String Models and Moduli Stabilisation}},  {\em Mod.
  Phys. Lett. A} {\bf 30} (2015), no.~07 1530004,
  [\href{http://arxiv.org/abs/1404.5151}{{\tt arXiv:1404.5151}}].

\bibitem{Slatyer:2016qyl}
T.~R. Slatyer and C.-L. Wu, {\it {General Constraints on Dark Matter Decay from
  the Cosmic Microwave Background}},  {\em Phys. Rev. D} {\bf 95} (2017), no.~2
  023010, [\href{http://arxiv.org/abs/1610.06933}{{\tt arXiv:1610.06933}}].

\bibitem{Djouadi:2005gi}
A.~Djouadi, {\it {The Anatomy of electro-weak symmetry breaking. I: The Higgs
  boson in the standard model}},  {\em Phys. Rept.} {\bf 457} (2008) 1--216,
  [\href{http://arxiv.org/abs/hep-ph/0503172}{{\tt hep-ph/0503172}}].

\bibitem{Djouadi:2005gj}
A.~Djouadi, {\it {The Anatomy of electro-weak symmetry breaking. II. The Higgs
  bosons in the minimal supersymmetric model}},  {\em Phys. Rept.} {\bf 459}
  (2008) 1--241, [\href{http://arxiv.org/abs/hep-ph/0503173}{{\tt
  hep-ph/0503173}}].

\bibitem{Franceschini:2015kwy}
R.~Franceschini, G.~F. Giudice, J.~F. Kamenik, et~al., {\it {What is the
  $\gamma \gamma$ resonance at 750 GeV?}},  {\em JHEP} {\bf 03} (2016) 144,
  [\href{http://arxiv.org/abs/1512.04933}{{\tt arXiv:1512.04933}}].

\bibitem{Takahashi:2020bpq}
F.~Takahashi, M.~Yamada, and W.~Yin, {\it {XENON1T Excess from Anomaly-Free
  Axionlike Dark Matter and Its Implications for Stellar Cooling Anomaly}},
  {\em Phys. Rev. Lett.} {\bf 125} (2020), no.~16 161801,
  [\href{http://arxiv.org/abs/2006.10035}{{\tt arXiv:2006.10035}}].

\bibitem{Han:2020dwo}
C.~Han, M.~L. L\'opez-Ib\'a\~nez, A.~Melis, O.~Vives, and J.~M. Yang, {\it
  {Anomaly-free leptophilic axionlike particle and its flavor violating
  tests}},  {\em Phys. Rev. D} {\bf 103} (2021), no.~3 035028,
  [\href{http://arxiv.org/abs/2007.08834}{{\tt arXiv:2007.08834}}].

\bibitem{Han:2022iig}
C.~Han, M.~L. L\'opez-Ib\'a\~nez, A.~Melis, O.~Vives, and J.~M. Yang, {\it
  {Anomaly-free ALP from non-Abelian flavor symmetry}},  {\em JHEP} {\bf 08}
  (2022) 306, [\href{http://arxiv.org/abs/2203.16376}{{\tt arXiv:2203.16376}}].

\bibitem{Sakurai:2022roq}
K.~Sakurai and F.~Takahashi, {\it {Anomaly-free axion dark matter in three
  Higgs doublet model and its phenomenological implications}},  {\em JHEP} {\bf
  07} (2022) 124, [\href{http://arxiv.org/abs/2203.17212}{{\tt
  arXiv:2203.17212}}]. [Erratum: JHEP 09, 132 (2023)].

\bibitem{Leutwyler:1989tn}
H.~Leutwyler and M.~A. Shifman, {\it {GOLDSTONE BOSONS GENERATE PECULIAR
  CONFORMAL ANOMALIES}},  {\em Phys. Lett. B} {\bf 221} (1989) 384--388.

\bibitem{Flambaum:2024zyt}
V.~V. Flambaum and I.~B. Samsonov, {\it {Limits on scalar dark matter
  interactions with particles other than the photon via loop corrections to the
  scalar-photon coupling}},  {\em Phys. Rev. D} {\bf 110} (2024), no.~7 075044,
  [\href{http://arxiv.org/abs/2403.02685}{{\tt arXiv:2403.02685}}].

\bibitem{Arias-Aragon:2020shv}
F.~Arias-Arag\'on, F.~D'Eramo, R.~Z. Ferreira, L.~Merlo, and A.~Notari, {\it
  {Production of Thermal Axions across the ElectroWeak Phase Transition}},
  {\em JCAP} {\bf 03} (2021) 090, [\href{http://arxiv.org/abs/2012.04736}{{\tt
  arXiv:2012.04736}}].

\bibitem{Bolliet:2020ofj}
B.~Bolliet, J.~Chluba, and R.~Battye, {\it {Spectral distortion constraints on
  photon injection from low-mass decaying particles}},  {\em Mon. Not. Roy.
  Astron. Soc.} {\bf 507} (2021), no.~3 3148--3178,
  [\href{http://arxiv.org/abs/2012.07292}{{\tt arXiv:2012.07292}}].

\bibitem{Cirelli:2012ut}
M.~Cirelli, E.~Moulin, P.~Panci, P.~D. Serpico, and A.~Viana, {\it {Gamma ray
  constraints on Decaying Dark Matter}},  {\em Phys. Rev. D} {\bf 86} (2012)
  083506, [\href{http://arxiv.org/abs/1205.5283}{{\tt arXiv:1205.5283}}].

\bibitem{Laha:2020ivk}
R.~Laha, J.~B. Mu\~noz, and T.~R. Slatyer, {\it {INTEGRAL constraints on
  primordial black holes and particle dark matter}},  {\em Phys. Rev. D} {\bf
  101} (2020), no.~12 123514, [\href{http://arxiv.org/abs/2004.00627}{{\tt
  arXiv:2004.00627}}].

\bibitem{Watson:2011dw}
C.~R. Watson, Z.-Y. Li, and N.~K. Polley, {\it {Constraining Sterile Neutrino
  Warm Dark Matter with Chandra Observations of the Andromeda Galaxy}},  {\em
  JCAP} {\bf 03} (2012) 018, [\href{http://arxiv.org/abs/1111.4217}{{\tt
  arXiv:1111.4217}}].

\bibitem{Horiuchi:2013noa}
S.~Horiuchi, P.~J. Humphrey, J.~Onorbe, et~al., {\it {Sterile neutrino dark
  matter bounds from galaxies of the Local Group}},  {\em Phys. Rev. D} {\bf
  89} (2014), no.~2 025017, [\href{http://arxiv.org/abs/1311.0282}{{\tt
  arXiv:1311.0282}}].

\bibitem{Foster:2022ajl}
J.~W. Foster, S.~Kumar, B.~R. Safdi, and Y.~Soreq, {\it {Dark Grand Unification
  in the axiverse: decaying axion dark matter and spontaneous baryogenesis}},
  {\em JHEP} {\bf 12} (2022) 119, [\href{http://arxiv.org/abs/2208.10504}{{\tt
  arXiv:2208.10504}}].

\bibitem{Perez:2016tcq}
K.~Perez, K.~C.~Y. Ng, J.~F. Beacom, et~al., {\it {Almost closing the
  \ensuremath{\nu}MSM sterile neutrino dark matter window with NuSTAR}},  {\em
  Phys. Rev. D} {\bf 95} (2017), no.~12 123002,
  [\href{http://arxiv.org/abs/1609.00667}{{\tt arXiv:1609.00667}}].

\bibitem{Roach:2019ctw}
B.~M. Roach, K.~C.~Y. Ng, K.~Perez, et~al., {\it {NuSTAR Tests of
  Sterile-Neutrino Dark Matter: New Galactic Bulge Observations and Combined
  Impact}},  {\em Phys. Rev. D} {\bf 101} (2020), no.~10 103011,
  [\href{http://arxiv.org/abs/1908.09037}{{\tt arXiv:1908.09037}}].

\bibitem{Ng:2019gch}
K.~C.~Y. Ng, B.~M. Roach, K.~Perez, et~al., {\it {New Constraints on Sterile
  Neutrino Dark Matter from $NuSTAR$ M31 Observations}},  {\em Phys. Rev. D}
  {\bf 99} (2019) 083005, [\href{http://arxiv.org/abs/1901.01262}{{\tt
  arXiv:1901.01262}}].

\bibitem{Roach:2022lgo}
B.~M. Roach, S.~Rossland, K.~C.~Y. Ng, et~al., {\it {Long-exposure NuSTAR
  constraints on decaying dark matter in the Galactic halo}},  {\em Phys. Rev.
  D} {\bf 107} (2023), no.~2 023009,
  [\href{http://arxiv.org/abs/2207.04572}{{\tt arXiv:2207.04572}}].

\bibitem{Coogan:2021rez}
A.~Coogan et~al., {\it {Hunting for dark matter and new physics with GECCO}},
  {\em Phys. Rev. D} {\bf 107} (2023), no.~2 023022,
  [\href{http://arxiv.org/abs/2101.10370}{{\tt arXiv:2101.10370}}].

\bibitem{Thorpe-Morgan:2020rwc}
C.~Thorpe-Morgan, D.~Malyshev, A.~Santangelo, et~al., {\it {THESEUS insights
  into axionlike particles, dark photon, and sterile neutrino dark matter}},
  {\em Phys. Rev. D} {\bf 102} (2020), no.~12 123003,
  [\href{http://arxiv.org/abs/2008.08306}{{\tt arXiv:2008.08306}}].

\bibitem{Neronov:2015kca}
A.~Neronov and D.~Malyshev, {\it {Toward a full test of the $\nu$MSM sterile
  neutrino dark matter model with Athena}},  {\em Phys. Rev. D} {\bf 93}
  (2016), no.~6 063518, [\href{http://arxiv.org/abs/1509.02758}{{\tt
  arXiv:1509.02758}}].

\bibitem{Dekker:2021bos}
A.~Dekker, E.~Peerbooms, F.~Zimmer, K.~C.~Y. Ng, and S.~Ando, {\it {Searches
  for sterile neutrinos and axionlike particles from the Galactic halo with
  eROSITA}},  {\em Phys. Rev. D} {\bf 104} (2021), no.~2 023021,
  [\href{http://arxiv.org/abs/2103.13241}{{\tt arXiv:2103.13241}}].

\bibitem{Ando:2021fhj}
S.~Ando et~al., {\it {Decaying dark matter in dwarf spheroidal galaxies:
  Prospects for x-ray and gamma-ray telescopes}},  {\em Phys. Rev. D} {\bf 104}
  (2021), no.~2 023022, [\href{http://arxiv.org/abs/2103.13242}{{\tt
  arXiv:2103.13242}}].

\bibitem{Todarello:2023hdk}
E.~Todarello, M.~Regis, J.~Reynoso-Cordova, et~al., {\it {Robust bounds on ALP
  dark matter from dwarf spheroidal galaxies in the optical MUSE-Faint
  survey}},  {\em JCAP} {\bf 05} (2024) 043,
  [\href{http://arxiv.org/abs/2307.07403}{{\tt arXiv:2307.07403}}].

\bibitem{Todarello:2024qci}
E.~Todarello and M.~Regis, {\it {Bounds on Axions-Like Particles Shining in the
  Ultra-Violet}},  \href{http://arxiv.org/abs/2412.02543}{{\tt
  arXiv:2412.02543}}.

\bibitem{Sun:2023acy}
Y.~Sun, J.~W. Foster, H.~Liu, J.~B. Mu\~noz, and T.~R. Slatyer, {\it
  {Inhomogeneous energy injection in the 21-cm power spectrum: Sensitivity to
  dark matter decay}},  {\em Phys. Rev. D} {\bf 111} (2025), no.~4 043015,
  [\href{http://arxiv.org/abs/2312.11608}{{\tt arXiv:2312.11608}}].

\bibitem{Ziegler:2023aoe}
R.~Ziegler, {\it {Flavor Probes of Axion Dark Matter}},  {\em PoS} {\bf
  DISCRETE2022} (2024) 086, [\href{http://arxiv.org/abs/2303.13353}{{\tt
  arXiv:2303.13353}}].

\bibitem{NA62:2021zjw}
{\bf NA62} Collaboration, E.~Cortina~Gil et~al., {\it {Measurement of the very
  rare $K^{+}\to {\pi}^{+}\nu \overline{\nu} $ decay}},  {\em
  JHEP} {\bf 06} (2021) 093, [\href{http://arxiv.org/abs/2103.15389}{{\tt
  arXiv:2103.15389}}].

\bibitem{Goudzovski:2022vbt}
E.~Goudzovski et~al., {\it {New physics searches at kaon and hyperon
  factories}},  {\em Rept. Prog. Phys.} {\bf 86} (2023), no.~1 016201,
  [\href{http://arxiv.org/abs/2201.07805}{{\tt arXiv:2201.07805}}].

\bibitem{Carrasco:2016kpy}
N.~Carrasco, P.~Lami, V.~Lubicz, et~al., {\it {$K \to \pi$ semileptonic form
  factors with $N_f=2+1+1$ twisted mass fermions}},  {\em Phys. Rev. D} {\bf
  93} (2016), no.~11 114512, [\href{http://arxiv.org/abs/1602.04113}{{\tt
  arXiv:1602.04113}}].

\bibitem{FermilabLattice:2018zqv}
{\bf Fermilab Lattice, MILC} Collaboration, A.~Bazavov et~al., {\it {$|V_{us}|$
  from $K_{\ell 3}$ decay and four-flavor lattice QCD}},  {\em Phys. Rev. D}
  {\bf 99} (2019), no.~11 114509, [\href{http://arxiv.org/abs/1809.02827}{{\tt
  arXiv:1809.02827}}].

\bibitem{FlavourLatticeAveragingGroupFLAG:2021npn}
{\bf Flavour Lattice Averaging Group (FLAG)} Collaboration, Y.~Aoki et~al.,
  {\it {FLAG Review 2021}},  {\em Eur. Phys. J. C} {\bf 82} (2022), no.~10 869,
  [\href{http://arxiv.org/abs/2111.09849}{{\tt arXiv:2111.09849}}].

\bibitem{MartinCamalich:2020dfe}
J.~Martin~Camalich, M.~Pospelov, P.~N.~H. Vuong, R.~Ziegler, and J.~Zupan, {\it
  {Quark Flavor Phenomenology of the QCD Axion}},  {\em Phys. Rev. D} {\bf 102}
  (2020), no.~1 015023, [\href{http://arxiv.org/abs/2002.04623}{{\tt
  arXiv:2002.04623}}].

\bibitem{Calibbi:2020jvd}
L.~Calibbi, D.~Redigolo, R.~Ziegler, and J.~Zupan, {\it {Looking forward to
  lepton-flavor-violating ALPs}},  {\em JHEP} {\bf 09} (2021) 173,
  [\href{http://arxiv.org/abs/2006.04795}{{\tt arXiv:2006.04795}}].

\bibitem{Jodidio:1986mz}
A.~Jodidio et~al., {\it {Search for Right-Handed Currents in Muon Decay}},
  {\em Phys. Rev. D} {\bf 34} (1986) 1967. [Erratum: Phys.Rev.D 37, 237
  (1988)].

\bibitem{TWIST:2014ymv}
{\bf TWIST} Collaboration, R.~Bayes et~al., {\it {Search for two body muon
  decay signals}},  {\em Phys. Rev. D} {\bf 91} (2015), no.~5 052020,
  [\href{http://arxiv.org/abs/1409.0638}{{\tt arXiv:1409.0638}}].

\bibitem{Jho:2022snj}
Y.~Jho, S.~Knapen, and D.~Redigolo, {\it {Lepton-flavor violating axions at MEG
  II}},  {\em JHEP} {\bf 10} (2022) 029,
  [\href{http://arxiv.org/abs/2203.11222}{{\tt arXiv:2203.11222}}].

\bibitem{Knapen:2023zgi}
S.~Knapen, K.~Langhoff, T.~Opferkuch, and D.~Redigolo, {\it {A Robust Search
  for Lepton Flavour Violating Axions at Mu3e}},
  \href{http://arxiv.org/abs/2311.17915}{{\tt arXiv:2311.17915}}.

\bibitem{Hill:2023dym}
R.~J. Hill, R.~Plestid, and J.~Zupan, {\it {Searching for new physics at
  $\mu\to e$ facilities with $\mu^+$ and
  $\pi^+$ decays at rest}},  {\em Phys. Rev. D} {\bf 109} (2024),
  no.~3 035025, [\href{http://arxiv.org/abs/2310.00043}{{\tt
  arXiv:2310.00043}}].

\bibitem{Belle-II:2022heu}
{\bf Belle-II} Collaboration, I.~Adachi et~al., {\it {Search for
  Lepton-Flavor-Violating \ensuremath{\tau} Decays to a Lepton and an Invisible
  Boson at Belle II}},  {\em Phys. Rev. Lett.} {\bf 130} (2023), no.~18 181803,
  [\href{http://arxiv.org/abs/2212.03634}{{\tt arXiv:2212.03634}}].

\bibitem{Belle-II:2023esi}
{\bf Belle-II} Collaboration, I.~Adachi et~al., {\it {Evidence for
  $B^+\to K^+\ensuremath{\nu}\ensuremath{\nu}$
  decays}},  {\em Phys. Rev. D} {\bf 109} (2024), no.~11 112006,
  [\href{http://arxiv.org/abs/2311.14647}{{\tt arXiv:2311.14647}}].

\bibitem{OHare:2020wah}
C.~A.~J. O'Hare and E.~Vitagliano, {\it {Cornering the axion with
  $CP$-violating interactions}},  {\em Phys. Rev. D} {\bf 102} (2020), no.~11
  115026, [\href{http://arxiv.org/abs/2010.03889}{{\tt arXiv:2010.03889}}].

\bibitem{Chen:2014oda}
Y.~J. Chen, W.~K. Tham, D.~E. Krause, et~al., {\it {Stronger Limits on
  Hypothetical Yukawa Interactions in the 30\textendash{}8000 nm Range}},  {\em
  Phys. Rev. Lett.} {\bf 116} (2016), no.~22 221102,
  [\href{http://arxiv.org/abs/1410.7267}{{\tt arXiv:1410.7267}}].

\bibitem{Kapner:2006si}
D.~J. Kapner, T.~S. Cook, E.~G. Adelberger, et~al., {\it {Tests of the
  gravitational inverse-square law below the dark-energy length scale}},  {\em
  Phys. Rev. Lett.} {\bf 98} (2007) 021101,
  [\href{http://arxiv.org/abs/hep-ph/0611184}{{\tt hep-ph/0611184}}].

\bibitem{Lee:2020zjt}
J.~G. Lee, E.~G. Adelberger, T.~S. Cook, S.~M. Fleischer, and B.~R. Heckel,
  {\it {New Test of the Gravitational $1/r^2$ Law at Separations down to 52
  $\mu$m}},  {\em Phys. Rev. Lett.} {\bf 124} (2020), no.~10 101101,
  [\href{http://arxiv.org/abs/2002.11761}{{\tt arXiv:2002.11761}}].

\bibitem{Yang:2012zzb}
S.-Q. Yang, B.-F. Zhan, Q.-L. Wang, et~al., {\it {Test of the Gravitational
  Inverse Square Law at Millimeter Ranges}},  {\em Phys. Rev. Lett.} {\bf 108}
  (2012) 081101.

\bibitem{Tan:2020vpf}
W.-H. Tan et~al., {\it {Improvement for Testing the Gravitational
  Inverse-Square Law at the Submillimeter Range}},  {\em Phys. Rev. Lett.} {\bf
  124} (2020), no.~5 051301.

\bibitem{Tu:2007zz}
L.-C. Tu, S.-G. Guan, J.~Luo, C.-G. Shao, and L.-X. Liu, {\it {Null Test of
  Newtonian Inverse-Square Law at Submillimeter Range with a Dual-Modulation
  Torsion Pendulum}},  {\em Phys. Rev. Lett.} {\bf 98} (2007) 201101.

\bibitem{Tan:2016vwu}
W.-H. Tan, S.-Q. Yang, C.-G. Shao, et~al., {\it {New Test of the Gravitational
  Inverse-Square Law at the Submillimeter Range with Dual Modulation and
  Compensation}},  {\em Phys. Rev. Lett.} {\bf 116} (2016), no.~13 131101.

\bibitem{Shifman:1978zn}
M.~A. Shifman, A.~I. Vainshtein, and V.~I. Zakharov, {\it {Remarks on Higgs
  Boson Interactions with Nucleons}},  {\em Phys. Lett. B} {\bf 78} (1978)
  443--446.

\bibitem{Cheng:2012qr}
H.-Y. Cheng and C.-W. Chiang, {\it {Revisiting Scalar and Pseudoscalar
  Couplings with Nucleons}},  {\em JHEP} {\bf 07} (2012) 009,
  [\href{http://arxiv.org/abs/1202.1292}{{\tt arXiv:1202.1292}}].

\bibitem{Raffelt:1996wa}
G.~G. Raffelt, {\em {Stars as laboratories for fundamental physics}: {The
  astrophysics of neutrinos, axions, and other weakly interacting particles}}.
\newblock 5, 1996.

\bibitem{Bottaro:2023gep}
S.~Bottaro, A.~Caputo, G.~Raffelt, and E.~Vitagliano, {\it {Stellar limits on
  scalars from electron-nucleus bremsstrahlung}},  {\em JCAP} {\bf 07} (2023)
  071, [\href{http://arxiv.org/abs/2303.00778}{{\tt arXiv:2303.00778}}].

\bibitem{Yamamoto:2023zlu}
Y.~Yamamoto and K.~Yoshioka, {\it {Stellar cooling limits on light scalar boson
  revisited}},  {\em Phys. Lett. B} {\bf 843} (2023) 138027,
  [\href{http://arxiv.org/abs/2303.03123}{{\tt arXiv:2303.03123}}].

\bibitem{Grifols:1986fc}
J.~A. Grifols and E.~Masso, {\it {Constraints on Finite Range Baryonic and
  Leptonic Forces From Stellar Evolution}},  {\em Phys. Lett. B} {\bf 173}
  (1986) 237--240.

\bibitem{Hardy:2016kme}
E.~Hardy and R.~Lasenby, {\it {Stellar cooling bounds on new light particles:
  plasma mixing effects}},  {\em JHEP} {\bf 02} (2017) 033,
  [\href{http://arxiv.org/abs/1611.05852}{{\tt arXiv:1611.05852}}].

\bibitem{CAST:2017uph}
{\bf CAST} Collaboration, V.~Anastassopoulos et~al., {\it {New CAST Limit on
  the Axion-Photon Interaction}},  {\em Nature Phys.} {\bf 13} (2017) 584--590,
  [\href{http://arxiv.org/abs/1705.02290}{{\tt arXiv:1705.02290}}].

\bibitem{Ellis:1982tk}
J.~R. Ellis, S.~Ferrara, and D.~V. Nanopoulos, {\it {CP Violation and
  Supersymmetry}},  {\em Phys. Lett. B} {\bf 114} (1982) 231--234.

\bibitem{Giudice:1998bp}
G.~F. Giudice and R.~Rattazzi, {\it {Theories with gauge mediated supersymmetry
  breaking}},  {\em Phys. Rept.} {\bf 322} (1999) 419--499,
  [\href{http://arxiv.org/abs/hep-ph/9801271}{{\tt hep-ph/9801271}}].

\bibitem{Randall:1998uk}
L.~Randall and R.~Sundrum, {\it {Out of this world supersymmetry breaking}},
  {\em Nucl. Phys. B} {\bf 557} (1999) 79--118,
  [\href{http://arxiv.org/abs/hep-th/9810155}{{\tt hep-th/9810155}}].

\bibitem{Giudice:1998xp}
G.~F. Giudice, M.~A. Luty, H.~Murayama, and R.~Rattazzi, {\it {Gaugino mass
  without singlets}},  {\em JHEP} {\bf 12} (1998) 027,
  [\href{http://arxiv.org/abs/hep-ph/9810442}{{\tt hep-ph/9810442}}].

\bibitem{Rattazzi:1999qg}
R.~Rattazzi, A.~Strumia, and J.~D. Wells, {\it {Phenomenology of deflected
  anomaly mediation}},  {\em Nucl. Phys. B} {\bf 576} (2000) 3--28,
  [\href{http://arxiv.org/abs/hep-ph/9912390}{{\tt hep-ph/9912390}}].

\bibitem{Enguita:2024nuq}
V.~Enguita, B.~Gavela, B.~Grinstein, and P.~Quilez, {\it {ALP contribution to
  the strong CP problem}},  {\em Phys. Rev. D} {\bf 110} (2024), no.~1 015024,
  [\href{http://arxiv.org/abs/2403.12133}{{\tt arXiv:2403.12133}}].

\bibitem{Pospelov:1999mv}
M.~Pospelov and A.~Ritz, {\it {Theta vacua, QCD sum rules, and the neutron
  electric dipole moment}},  {\em Nucl. Phys. B} {\bf 573} (2000) 177--200,
  [\href{http://arxiv.org/abs/hep-ph/9908508}{{\tt hep-ph/9908508}}].

\bibitem{Abel:2020pzs}
C.~Abel et~al., {\it {Measurement of the Permanent Electric Dipole Moment of
  the Neutron}},  {\em Phys. Rev. Lett.} {\bf 124} (2020), no.~8 081803,
  [\href{http://arxiv.org/abs/2001.11966}{{\tt arXiv:2001.11966}}].

\bibitem{Arias:2012az}
P.~Arias, D.~Cadamuro, M.~Goodsell, et~al., {\it {WISPy Cold Dark Matter}},
  {\em JCAP} {\bf 06} (2012) 013, [\href{http://arxiv.org/abs/1201.5902}{{\tt
  arXiv:1201.5902}}].

\bibitem{Blinov:2019rhb}
N.~Blinov, M.~J. Dolan, P.~Draper, and J.~Kozaczuk, {\it {Dark matter targets
  for axionlike particle searches}},  {\em Phys. Rev. D} {\bf 100} (2019),
  no.~1 015049, [\href{http://arxiv.org/abs/1905.06952}{{\tt
  arXiv:1905.06952}}].

\bibitem{OHare:2024nmr}
C.~A.~J. O'Hare, {\it {Cosmology of axion dark matter}},  {\em PoS} {\bf
  COSMICWISPers} (2024) 040, [\href{http://arxiv.org/abs/2403.17697}{{\tt
  arXiv:2403.17697}}].

\bibitem{Visinelli:2009kt}
L.~Visinelli and P.~Gondolo, {\it {Axion cold dark matter in non-standard
  cosmologies}},  {\em Phys. Rev. D} {\bf 81} (2010) 063508,
  [\href{http://arxiv.org/abs/0912.0015}{{\tt arXiv:0912.0015}}].

\bibitem{Kawasaki:2000en}
M.~Kawasaki, K.~Kohri, and N.~Sugiyama, {\it {MeV scale reheating temperature
  and thermalization of neutrino background}},  {\em Phys. Rev. D} {\bf 62}
  (2000) 023506, [\href{http://arxiv.org/abs/astro-ph/0002127}{{\tt
  astro-ph/0002127}}].

\bibitem{Hannestad:2004px}
S.~Hannestad, {\it {What is the lowest possible reheating temperature?}},  {\em
  Phys. Rev. D} {\bf 70} (2004) 043506,
  [\href{http://arxiv.org/abs/astro-ph/0403291}{{\tt astro-ph/0403291}}].

\bibitem{DEramo:2017ecx}
F.~D'Eramo, N.~Fernandez, and S.~Profumo, {\it {Dark Matter Freeze-in
  Production in Fast-Expanding Universes}},  {\em JCAP} {\bf 02} (2018) 046,
  [\href{http://arxiv.org/abs/1712.07453}{{\tt arXiv:1712.07453}}].

\bibitem{Badziak:2024szg}
M.~Badziak, K.~Harigaya, M.~\L{}ukawski, and R.~Ziegler, {\it {Thermal
  production of astrophobic axions}},  {\em JHEP} {\bf 09} (2024) 136,
  [\href{http://arxiv.org/abs/2403.05621}{{\tt arXiv:2403.05621}}].

\bibitem{Cadamuro:2010cz}
D.~Cadamuro, S.~Hannestad, G.~Raffelt, and J.~Redondo, {\it {Cosmological
  bounds on sub-MeV mass axions}},  {\em JCAP} {\bf 02} (2011) 003,
  [\href{http://arxiv.org/abs/1011.3694}{{\tt arXiv:1011.3694}}].

\bibitem{Elahi:2014fsa}
F.~Elahi, C.~Kolda, and J.~Unwin, {\it {UltraViolet Freeze-in}},  {\em JHEP}
  {\bf 03} (2015) 048, [\href{http://arxiv.org/abs/1410.6157}{{\tt
  arXiv:1410.6157}}].

\bibitem{Silva-Malpartida:2024emu}
J.~Silva-Malpartida, N.~Bernal, J.~Jones-P\'erez, and R.~A. Lineros, {\it {From
  WIMPs to FIMPs: Impact of Early Matter Domination}},
  \href{http://arxiv.org/abs/2408.08950}{{\tt arXiv:2408.08950}}.

\bibitem{Viel:2004bf}
M.~Viel, M.~G. Haehnelt, and V.~Springel, {\it {Inferring the dark matter power
  spectrum from the Lyman-alpha forest in high-resolution QSO absorption
  spectra}},  {\em Mon. Not. Roy. Astron. Soc.} {\bf 354} (2004) 684,
  [\href{http://arxiv.org/abs/astro-ph/0404600}{{\tt astro-ph/0404600}}].

\bibitem{Boyarsky:2008xj}
A.~Boyarsky, J.~Lesgourgues, O.~Ruchayskiy, and M.~Viel, {\it {Lyman-alpha
  constraints on warm and on warm-plus-cold dark matter models}},  {\em JCAP}
  {\bf 05} (2009) 012, [\href{http://arxiv.org/abs/0812.0010}{{\tt
  arXiv:0812.0010}}].

\bibitem{Viel:2013fqw}
M.~Viel, G.~D. Becker, J.~S. Bolton, and M.~G. Haehnelt, {\it {Warm dark matter
  as a solution to the small scale crisis: New constraints from high redshift
  Lyman-\ensuremath{\alpha} forest data}},  {\em Phys. Rev. D} {\bf 88} (2013)
  043502, [\href{http://arxiv.org/abs/1306.2314}{{\tt arXiv:1306.2314}}].

\bibitem{Baur:2015jsy}
J.~Baur, N.~Palanque-Delabrouille, C.~Y\`eche, C.~Magneville, and M.~Viel, {\it
  {Lyman-alpha Forests cool Warm Dark Matter}},  {\em JCAP} {\bf 08} (2016)
  012, [\href{http://arxiv.org/abs/1512.01981}{{\tt arXiv:1512.01981}}].

\bibitem{Irsic:2017ixq}
V.~Ir\v{s}i\v{c} et~al., {\it {New Constraints on the free-streaming of warm
  dark matter from intermediate and small scale Lyman-$\alpha$ forest data}},
  {\em Phys. Rev. D} {\bf 96} (2017), no.~2 023522,
  [\href{http://arxiv.org/abs/1702.01764}{{\tt arXiv:1702.01764}}].

\bibitem{DEramo:2020gpr}
F.~D'Eramo and A.~Lenoci, {\it {Lower mass bounds on FIMP dark matter produced
  via freeze-in}},  {\em JCAP} {\bf 10} (2021) 045,
  [\href{http://arxiv.org/abs/2012.01446}{{\tt arXiv:2012.01446}}].

\bibitem{Ballesteros:2020adh}
G.~Ballesteros, M.~A.~G. Garcia, and M.~Pierre, {\it {How warm are non-thermal
  relics? Lyman-$\alpha$ bounds on out-of-equilibrium dark matter}},  {\em
  JCAP} {\bf 03} (2021) 101, [\href{http://arxiv.org/abs/2011.13458}{{\tt
  arXiv:2011.13458}}].

\bibitem{Decant:2021mhj}
Q.~Decant, J.~Heisig, D.~C. Hooper, and L.~Lopez-Honorez, {\it
  {Lyman-\ensuremath{\alpha} constraints on freeze-in and superWIMPs}},  {\em
  JCAP} {\bf 03} (2022) 041, [\href{http://arxiv.org/abs/2111.09321}{{\tt
  arXiv:2111.09321}}].

\bibitem{DEramo:2024jhn}
F.~D'Eramo and A.~Lenoci, {\it {Back to the phase space: Thermal axion dark
  radiation via couplings to standard model fermions}},  {\em Phys. Rev. D}
  {\bf 110} (2024), no.~11 116028, [\href{http://arxiv.org/abs/2410.21253}{{\tt
  arXiv:2410.21253}}].

\bibitem{Badziak:2024qjg}
M.~Badziak and M.~Laletin, {\it {Precise predictions for the QCD axion
  contribution to dark radiation with full phase-space evolution}},
  \href{http://arxiv.org/abs/2410.18186}{{\tt arXiv:2410.18186}}.

\bibitem{Antusch:2013jca}
S.~Antusch and V.~Maurer, {\it {Running quark and lepton parameters at various
  scales}},  {\em JHEP} {\bf 11} (2013) 115,
  [\href{http://arxiv.org/abs/1306.6879}{{\tt arXiv:1306.6879}}].

\bibitem{Ding:2020yen}
G.-J. Ding and F.~Feruglio, {\it {Testing Moduli and Flavon Dynamics with
  Neutrino Oscillations}},  {\em JHEP} {\bf 06} (2020) 134,
  [\href{http://arxiv.org/abs/2003.13448}{{\tt arXiv:2003.13448}}].

\bibitem{Ferrara:1989bc}
S.~Ferrara, D.~Lust, A.~D. Shapere, and S.~Theisen, {\it {Modular Invariance in
  Supersymmetric Field Theories}},  {\em Phys. Lett. B} {\bf 225} (1989) 363.

\bibitem{Ferrara:1989qb}
S.~Ferrara, .~D. Lust, and S.~Theisen, {\it {Target Space Modular Invariance
  and Low-Energy Couplings in Orbifold Compactifications}},  {\em Phys. Lett.
  B} {\bf 233} (1989) 147--152.

\bibitem{Feruglio:2017spp}
F.~Feruglio, {\em {Are neutrino masses modular forms?}}, pp.~227--266.
\newblock 2019.
\newblock [\href{http://arxiv.org/abs/1706.08749}{{\tt arXiv:1706.08749}}].

\bibitem{Baur:2019kwi}
A.~Baur, H.~P. Nilles, A.~Trautner, and P.~K.~S. Vaudrevange, {\it {Unification
  of Flavor, CP, and Modular Symmetries}},  {\em Phys. Lett. B} {\bf 795}
  (2019) 7--14, [\href{http://arxiv.org/abs/1901.03251}{{\tt
  arXiv:1901.03251}}].

\bibitem{Novichkov:2019sqv}
P.~P. Novichkov, J.~T. Penedo, S.~T. Petcov, and A.~V. Titov, {\it {Generalised
  CP Symmetry in Modular-Invariant Models of Flavour}},  {\em JHEP} {\bf 07}
  (2019) 165, [\href{http://arxiv.org/abs/1905.11970}{{\tt arXiv:1905.11970}}].

\bibitem{Baur:2019iai}
A.~Baur, H.~P. Nilles, A.~Trautner, and P.~K.~S. Vaudrevange, {\it {A String
  Theory of Flavor and $\mathscr {CP}$}},  {\em Nucl. Phys. B} {\bf 947} (2019)
  114737, [\href{http://arxiv.org/abs/1908.00805}{{\tt arXiv:1908.00805}}].

\bibitem{Wess:1992cp}
J.~Wess and J.~Bagger, {\em {Supersymmetry and supergravity}}.
\newblock Princeton University Press, Princeton, NJ, USA, 1992.

\bibitem{Feruglio:2021dte}
F.~Feruglio, V.~Gherardi, A.~Romanino, and A.~Titov, {\it {Modular invariant
  dynamics and fermion mass hierarchies around $\tau = i$}},  {\em JHEP} {\bf
  05} (2021) 242, [\href{http://arxiv.org/abs/2101.08718}{{\tt
  arXiv:2101.08718}}].

\bibitem{Novichkov:2021evw}
P.~P. Novichkov, J.~T. Penedo, and S.~T. Petcov, {\it {Fermion mass
  hierarchies, large lepton mixing and residual modular symmetries}},  {\em
  JHEP} {\bf 04} (2021) 206, [\href{http://arxiv.org/abs/2102.07488}{{\tt
  arXiv:2102.07488}}].

\bibitem{Feruglio:2022koo}
F.~Feruglio, {\it {Universal Predictions of Modular Invariant Flavor Models
  near the Self-Dual Point}},  {\em Phys. Rev. Lett.} {\bf 130} (2023), no.~10
  101801, [\href{http://arxiv.org/abs/2211.00659}{{\tt arXiv:2211.00659}}].

\bibitem{Feruglio:2023mii}
F.~Feruglio, {\it {Fermion masses, critical behavior and universality}},  {\em
  JHEP} {\bf 03} (2023) 236, [\href{http://arxiv.org/abs/2302.11580}{{\tt
  arXiv:2302.11580}}].

\bibitem{Ding:2024xhz}
G.-J. Ding, F.~Feruglio, and X.-G. Liu, {\it {Universal predictions of Siegel
  modular invariant theories near the fixed points}},  {\em JHEP} {\bf 05}
  (2024) 052, [\href{http://arxiv.org/abs/2402.14915}{{\tt arXiv:2402.14915}}].

\end{thebibliography}\endgroup

\end{document}